\documentclass[10pt,preprint2]{aastex}
\shorttitle{Possible planet candidate in NGC~2158}
\shortauthors{Mochejska et al.}
\begin{document}

\title{Planets in Stellar Clusters Extensive Search. IV.~A detection
of a possible transiting planet candidate in the open cluster
NGC 2158.\altaffilmark{1}} \altaffiltext{1}{Based on data from the
FLWO 1.2m telescope}

\author{B.~J.~Mochejska\altaffilmark{2}}

\affil{Purdue University, Department of Physics, 525 Northwestern
Ave., West Lafayette, IN~47907}
\email{bmochejs@cfa.harvard.edu}
\author{K.~Z.~Stanek}
\affil{Department of Astronomy, The Ohio State University,
140 W.\ 18th Avenue, Columbus, OH 43210}
\email{kstanek@astronomy.ohio-state.edu}
\author{D.~D.~Sasselov, A.~H.~Szentgyorgyi, E.~Adams,
R.~L.~Cooper, J.~B.~Foster, J.~D.~Hartman, R.~C.~Hickox, K.~Lai,
M.~Westover \& J.~N.~Winn\altaffilmark{2}}
\affil{Harvard-Smithsonian Center for Astrophysics, 60 Garden St.,
Cambridge, MA~02138}
\email{sasselov, saint@cfa.harvard.edu, era@improbable.org,
rcooper, jfoster, jhartman, rhickox, klai, mwestover,
jwinn@cfa.harvard.edu}
\altaffiltext{2}{Hubble Fellow}

\begin{abstract}
We have undertaken a long-term project, Planets in Stellar Clusters
Extensive Search (PISCES), to search for transiting planets in open
clusters. In this paper we present the results for NGC~2158, an
intermediate age, populous cluster. We have monitored the cluster for
over 260 hours, spread over 59 nights. We have detected one candidate
transiting low luminosity object, with eclipse depth of 3.7\% in the
$R$-band. If the host star is a member of the cluster, the eclipse
depth is consistent with a 1.7 $R_J$ object. Cluster membership of the
host is supported by its location on the cluster main sequence (MS)
and its close proximity to the cluster center (2\arcmin). We have
discovered two other stars exhibiting low-amplitude (4-5\%) transits,
V64 and V70, but they are most likely blends or field stars.  Given
the photometric precision and temporal coverage of our observations,
and current best estimates for the frequency and radii of short-period
planets, the expected number of detectable transiting planets in our
sample is 0.13. We have observed four outbursts for the candidate
cataclysmic variable V57. We have discovered 40 new variable stars in
the cluster, bringing the total number of identified variables to 97,
and present for them high precision light curves, spanning 13 months.
\end{abstract}


\section{{\sc Introduction}}
We have undertaken a long-term project, Planets in Stellar Clusters
Extensive Search (PISCES), to search for transiting planets in open
clusters. To date we have published a feasibility study based on one
season of data for NGC~6791 (Mochejska et al.\ 2002, hereafter
Paper~I) and a catalog of 57 variable stars for our second target,
NGC~2158, based on the data from the first observing season (Mochejska
et al.\ 2004, hereafter Paper~II). We have also published the results
of an extensive search for transiting planets in NGC 6791, based on
over 300 hours of observations, spread over 84 nights (Mochejska et
al.\ 2005, hereafter Paper III). We have not detected any promising
candidates, and derived an estimate of 1.7 expected transiting
planets.

In this paper we present the results of a search for transiting
planets in the open cluster NGC~2158 $[(\alpha,\delta)_{2000}=(6^h7^m,
+24^{\circ}0'); (l,b)=(186\fdg 63, +1\fdg 78)]$. It is a very
populous, intermediate age ($\tau$=2-3 Gyr), rather metal poor
([Fe/H]=-0.46) open cluster, located at a distance of 3.6 kpc (Carraro
et al.\ 2002, hereafter Ca02; Christian, Heasley and Janes 1985).

Searching for planets in open clusters eliminates the problem of false
detections due to blended eclipsing binary stars, which are a
significant contaminant in the Galactic field searches (over 90\% of
all candidates; Konacki et al.\ 2003; Udalski et al.\ 2002a, 2002b).
Blending causes a large decrease of the depth of the eclipses and
mimics the transit of a much smaller object, such as a planet. As
opposed to dense star fields in the disk of our Galaxy, open clusters
located away from the galactic plane are sparse enough for blending to
be negligible.

There are two key elements in a survey for transiting planets. The
most commonly emphasized requirement is the high photometric
precision, at the 1\% level. The more often overlooked factor is the
need for very extensive temporal coverage.

Extensive temporal coverage is important because even for planets with
periods between 1 and 2 days, the fractional transit length is only
$\sim$5\% of the period, and it drops to $\sim$2\% for periods 2-10
days. During the remaining 95-98\% of the period the system is
photometrically indistinguishable from stars without transiting
planets. To our best knowledge, PISCES is the most extensive search
for transiting planets in open clusters in terms of temporal coverage
with a 1 m telescope.

The paper is arranged as follows: \S 2 describes the observations, \S
3 summarizes the reduction procedure, \S 4 outlines the search
strategy for transiting planets, \S 5 gives an estimate of the
expected number of transiting planet detections and \S 6 contains the
variable star catalog. Concluding remarks are found in \S 7.

\section{{\sc Observations}} 
\begin{figure}[t]
\plotone{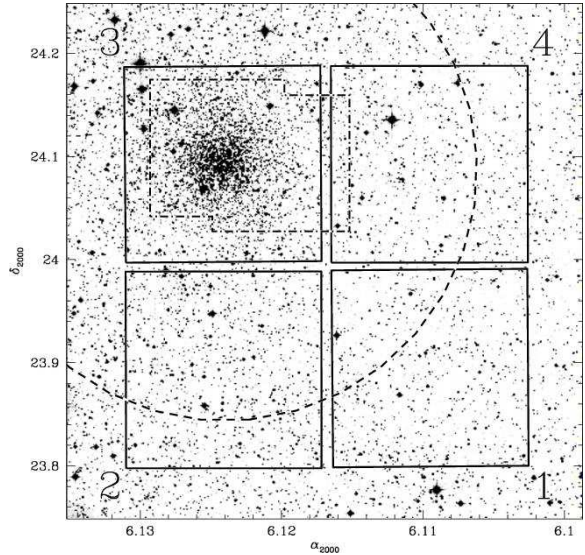}
\caption{Digital Sky Survey image of NGC~2158 showing the field of
view of the 4Shooter. The chips are numbered clockwise from 1 to 4
starting from the bottom right chip. NGC~2158 is centered on chip~3.
North is up and east is to the left.}
\label{chips}
\end{figure}

The data analyzed in this paper were obtained at the Fred Lawrence
Whipple Observatory (FLWO) 1.2 m telescope using the 4Shooter CCD
mosaic with four thinned, back side illuminated AR coated Loral
$2048^2$ CCDs (Szentgyorgyi et al.\ in preparation).  The camera, with
a pixel scale of $0\farcs 33$ pixel$^{-1}$, gives a field of view of
$11\farcm 4\times 11\farcm 4$ for each chip. The cluster was centered
on chip~3 (Fig.~\ref{chips}). The data were collected during 59
nights, from 2003 January 3 to 2004 February 17. A total of
$965\times 900$~s $R$ and $223\times 450$~s $V$-band exposures were
obtained. 

\section{{\sc Data Reduction}}

\subsection{{\it Image Subtraction Photometry}}
\label{imsub}
The preliminary processing of the CCD frames was performed with the
standard routines in the IRAF ccdproc package.\footnote{IRAF is
distributed by the National Optical Astronomy Observatories, which are   
operated by the Association of Universities for Research in Astronomy,
Inc., under cooperative agreement with the NSF.}

Photometry was extracted using the ISIS image subtraction package
(Alard \& Lupton 1998; Alard 2000), as described in detail in Papers~I
and III.

The ISIS reduction procedure consists of the following steps: (1)  
transformation of all frames to a common $(x,y)$ coordinate grid; (2)
construction of a reference image from several of the best exposures;    
(3) subtraction of each frame from the reference image; (4) selection 
of stars to be photometered and (5) extraction of profile photometry
from the subtracted images.

We used the same parameters for image subtraction as in Paper III.
The reference images were constructed from 25 best exposures in $R$
and 12 in $V$.

\begin{figure}[t]
\plotone{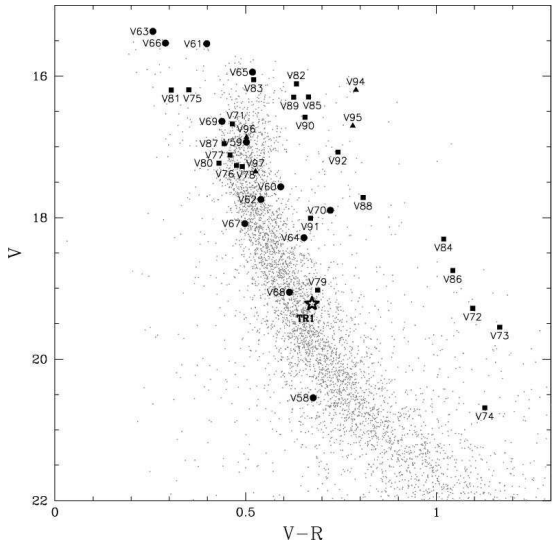}
\caption{$V/\vr$ CMD for chip~3, centered on NGC~2158. Newly
discovered eclipsing binaries are plotted with circles, other periodic
variables with squares and the non-periodic variables with triangles.
The low-luminosity transiting object candidate, TR1, is plotted with
a star. }
\label{fig:cmd}
\end{figure}

\begin{figure*}[t]
\epsscale{2.3}
\plottwo{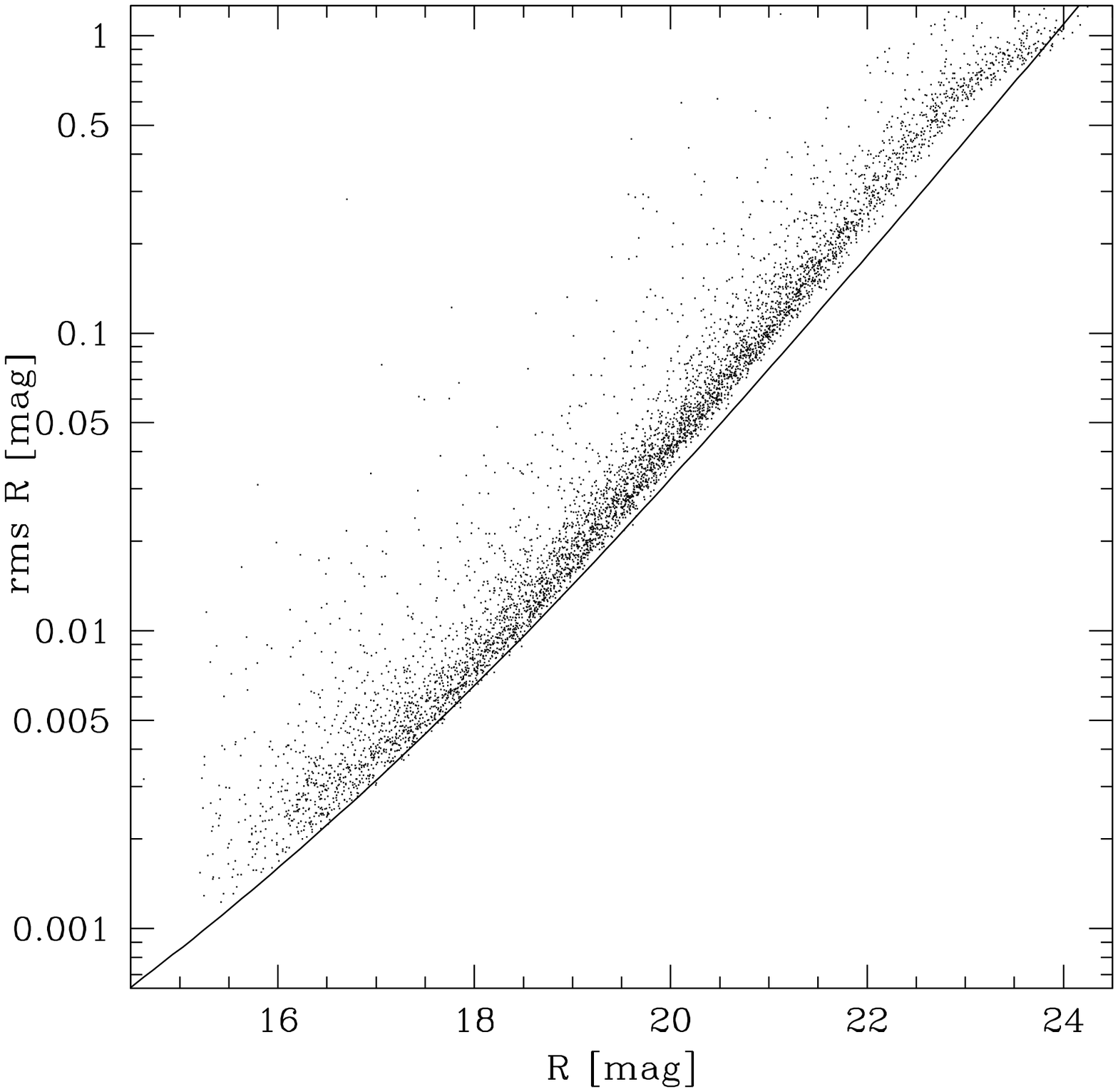}{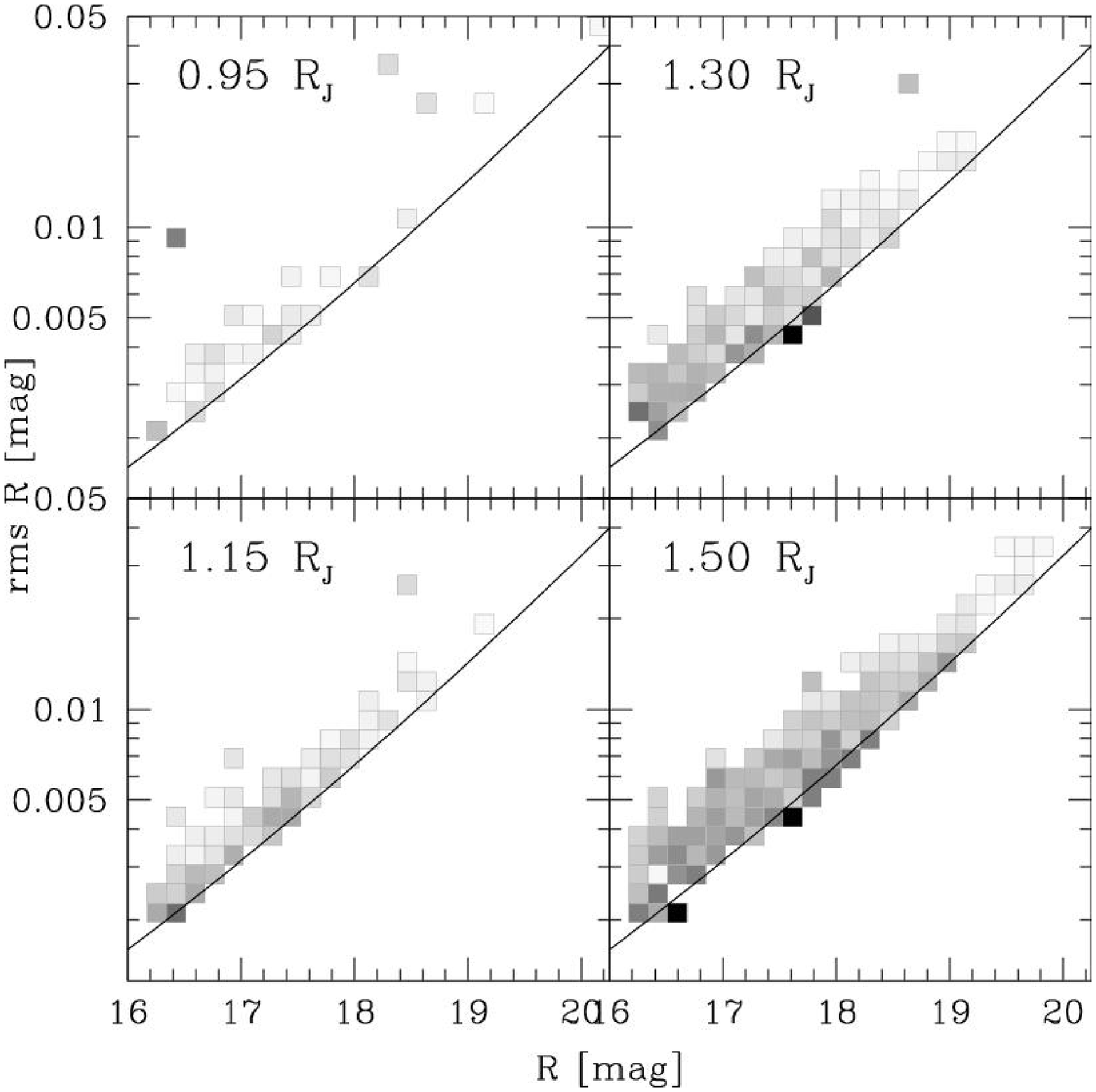}
\epsscale{1}
\caption{$Left$: The rms scatter of the $R$-band light curves for
stars on chip~3 with at least 500 data points. The solid curve
indicates the photometric precision limit due to Poisson noise of the
star and average sky brightness. $Right$: Detection efficiency
of 0.95, 1.15, 1.30 and 1.50 $R_J$ planets as a function of magnitude
and rms scatter (white: 0\%, black 100\%), determined in \S
\ref{sec:detef}.}
\label{fig:rms}
\end{figure*}

\subsection{{\it Calibration}}

To calibrate the $VR$ photometry we used two photometric catalogs: Ca02
and Kharchenko et al.\ 1997 (hereafter KAS97). The CCD photometry from
Ca02 has higher accuracy, while the photographic catalog from KAS97
overlaps all four chips, but it has a different zeropoint and larger
internal scatter, as described in Paper~II.

The $R$-band photometry was calibrated from the photometry published
by Ca02. We used 1685 stars above $R=18.5$ to determine the zero point
of the magnitude scale for chip 3. The rms scatter around the offset
was 0.09. Offsets between chip 3 and the remaining chips were
determined from an image, taken with chip 3, centered on the center of
the array. The number of stars used to determine the offset was 121,
135 and 293 for chips 1, 2 and 4, respectively.

As a consistency check, we compared these offsets with those derived
independently between chips 1, 2 and KAS97, and chip 4 and Ca02. The
differences in the offsets were 0.215, 0.181, and 0.010, based on 19,
90 and 36 stars, respectively. For chip 4, as expected, there is
excellent agreement, while for chips 1 and 2 the different zero point
of the two catalogs is apparent.

The $V$-band photometry for chips 1 and 2 was calibrated against
KAS97. To determine the zero points we used 19 and 90 stars above 17
mag for chips 1 and 2, respectively. The rms scatter of the residuals
was 0.13 on both chips.

The $V$-band photometry for chips 3 and 4 was calibrated against Ca02.
The zero points were derived from 1554 and 36 stars above 19 and 20
mag, and the rms scatter of the residuals was 0.09 and 0.06,
respectively.

Figure~\ref{fig:cmd} shows the calibrated V/V-R color-magnitude
diagram (CMD) for the chip~3 reference image.

\subsection{{\it Astrometry}}
Equatorial coordinates were determined for the $R$-band reference
image star lists. The transformation from rectangular to equatorial
coordinates was derived using 596, 710, 1952 and 615 transformation
stars from the 2MASS catalog (Cutri et al.\ 2003) in chips~1 through
4, respectively. The rms deviation between the catalog and the
computed coordinates for the transformation stars was $0\farcs 12$ in
right ascension and $0\farcs 11$ in declination.

\section{{\sc Search for Transiting Planets}}

\subsection{{\it Further Data Processing}}

We rejected from further analysis 96 $R$-band epochs where fewer than
4000 stars were detected on chip~3 by DAOphot (Stetson 1987). We also
rejected additional 29 to 31 bad quality images from four nights. For
chip 3, which suffers from moderate crowding and contains a significant
number of saturated stars, we also rejected additional 78 epochs where
the full-width at half-maximum (FWHM) of the stellar point-spread
function (PSF) was greater than 10 pixels, and the sky level was over
10,000 ADU. This left us with 807, 809, 729 and 809 highest quality
$R$-band exposures on chips~1-4, with a median seeing of $2\farcs
3$. We also removed 6 $V$-band images, which left us with 217
exposures with a median seeing of $2\farcs 6$.

In the NGC 6791 data, analyzed in Paper III, we noticed in the light
curves the presence of offsets between different runs. These were
probably due to the periodic UV flooding of the CCD camera, which
alters its quantum efficiency as a function of wavelength. In the NGC
2158 dataset analyzed here this problem was found to be much less
prominent, possibly due to the fact that this data set was obtained
over a shorter period of time (12.5 months, compared to 24 months for
NGC 6791). To prevent the transit detection algorithm from mistaking
these changes in brightness for transits, we corrected the light
curves using the method proposed by Tamuz et al.\ (2005). This
algorithm was originally envisioned to correct for color-dependent
atmospheric extinction, but can be used to correct for any linear
systematic effects. We solve for coefficients $c_i$ and $a_j$ that
minimize the following equation:
\begin{equation}
S_i^2 = \sum_{j}{\frac{(r_{ij} - c_i a_j)^2}{\sigma_{ij}^2}}
\end{equation}
where $r_{ij}$ is the residual for the observation of the $i$-th star
on the $j$-th image, or the star's average-subtracted magnitude, and
$\sigma_{ij}$ is the uncertainty of the measurement of star $i$ on
image $j$. Following Tamuz et al.\ (2005), the airmasses were used as
initial $a_j$ coefficients. We chose to run this algorithm four
times. After the fourth application, the amplitude in $a_j$ and
scatter in $c_i$ decreased about two times, relative to the first
run. As described in \S \ref{subsec:detef}, this cleaning procedure
significantly improves our detection efficiency.

The left panel of Fig.~\ref{fig:rms} shows the rms scatter of the
$R$-band light curves for stars on chip~3 with at least 500 data points.
The solid curve indicates the photometric precision limit due to Poisson
noise of the star and average sky brightness. The internal scatter of
the rms distribution in NGC 2158 is somewhat larger than in NGC 6791
(Fig.\ 3 in Paper III). This is mostly due to the fact that NGC 2158 is
a much more centrally concentrated cluster (it has 27\% more stars
within the central 1\arcmin\ than does NGC 6791). Most of the stars with
higher than average rms at a given magnitude are located at the cluster
center. Also, the average seeing for the NGC 2158 dataset was inferior
to the one for NGC 6791: $2\farcs 3$, compared to $2\farcs 1$.

The right panel of Fig.~\ref{fig:rms} shows the detection efficiency
of 0.95, 1.15, 1.30 and 1.50 $R_J$ planets as a function of magnitude
and rms scatter (white: 0\%, black 100\%), determined in \S
\ref{sec:detef}.

\subsection{{\it Selection of Transiting Planet Candidates}}
\label{sec:cand}
For further analysis we selected stars with at least 500 good epochs,
magnitudes $R>16.29$ (the main sequence turnoff; MSTO) and light
curve rms below 0.05 mag. This left us with 5159 stars (675, 981, 2680
and 823 stars on chips~1-4, respectively).

To select transiting planet candidates we used the box-fitting
least-squares (BLS) method (Kov{\' a}cs, Zucker, \& Mazeh 2002).
Adopting a cutoff of 6 in Signal Detection Efficiency (SDE) and 9 in
effective signal-to-noise ratio ($\alpha$), we selected 70 candidates:
3, 21, 17 and 28 on chips 1 through 4, respectively. Many of the
candidates turned out to be known eclipsing binaries, some had
sinusoidal light curves or periods which were multiples of a day.

It has been pointed out to us by G.\ B{\'a}kos (private communication)
that when the number of transits is small, SDE is not a good statistic.
We also examined additional 102 candidates with SDE$<6$ and $\alpha>20$
(13, 48, 6, 35 on chips 1-4, respectively). Most of the candidates had
very discrepant data on one or two nights, most likely due to a nearby
saturated star or an unidentified bad column.

\begin{figure}[t]
\plotone{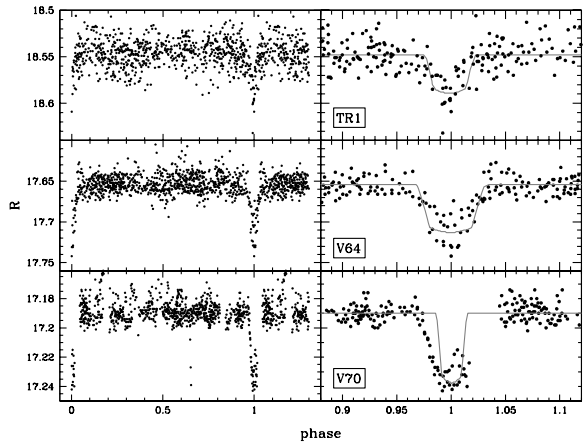}
\caption{$R$-band light curves for three transiting low luminosity
object candidates TR1 ($top$), V64 ($middle$) and V70 ($bottom$).
$Left$: entire phased light curves; $right$: a closeup of the eclipse,
with a simple model superimposed (solid line).}
\label{lc:tr3}
\end{figure}

\begin{figure}[t]
\plotone{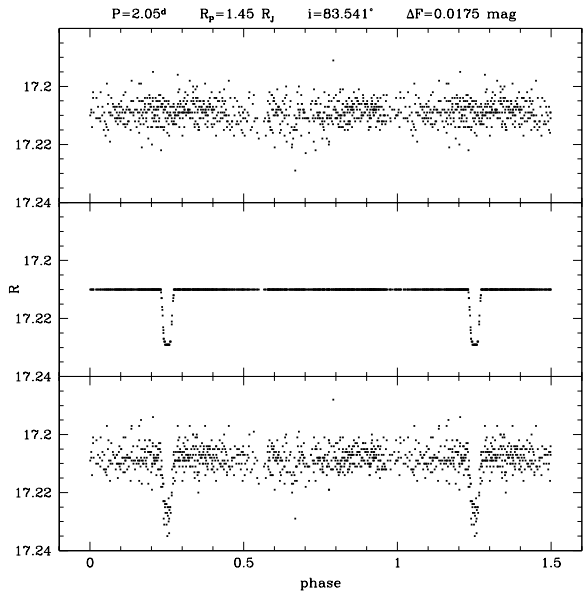}
\caption{Original ($top$), model ($middle$) and combined ($bottom$)
light curves for a star with $R=17.21$ and a planet with a period of
2.05 days, radius of 1.45 $R_J$ and inclination of 84$^{\circ}$.}
\label{fig:tr}
\end{figure}

\subsection{{\it Candidate Transiting Planets/Low Luminosity Objects}}

We identified one transiting object, TR1, with an $R$-band amplitude
of 3.7\% and $R$-band magnitude of 18.54. Its $R$-band light curve is
shown in the top left panel of Fig.~\ref{lc:tr3} and its parameters
are listed in Tab.~\ref{tab:tr}. Using the simple model described in
Paper III and \S \ref{sec:mod}, we find that the light curve is
consistent with a 0.87 $R_\odot$ star with a 1.66 $R_J$ low luminosity
companion, assuming an orbital inclination of 90$^\circ$. The top right
panel of Fig.~\ref{lc:tr3} shows this model (solid line) superimposed
on the $R$-band light curve. It is likely that the star is a cluster
member, as it is located $2\farcm 1$ from cluster center and on the
CMD it falls on the cluster MS. The only other possibility
is that it is a stellar binary with grazing eclipses.

The radii of the seven known Jupiter-mass (0.53-1.45 $M_J$) transiting
planets span the range from 1 to 1.32 $R_J$ (The Extrasolar Planets
Encyclopaedia
\footnote{See http://www.obspm.fr/planets}). An eighth planet, HD
149026b, recently found to display transits has a mass of 0.36 $M_J$
and a radius of 0.726 $R_J$ (Charbonneau et al.\ 2005). In light of
this, TR1 seems too large to be a planet. Based on models of brown
dwarfs (BD) and very low mass stars (VLM), at an age of 1 Gyr a radius
of 1.66 $R_J$ would correspond to a 0.085 $M_\odot$ star, and at 3.2
Gyr to a 0.140 $M_\odot$ star (Baraffe et al.\ 2001). Assuming an age
of 2 Gyr for NGC~2158, as determined by Ca02, such a radius would
correspond to $\sim$0.1 $M_\odot$ star. Precise determinations of
radii and masses for BDs and VLM stars are few. Until recently, only
the nearest such systems could be studied, due to their low
luminosities and small radii (hence shallow eclipses). The advent of
high precision mass photometry has stimulated interest in the search
for such systems (Hebb et al.\ 2004; Pont et al.\ 2005; Pinfield et
al.\ 2005).

Among the candidates with low $SDE$ we found two stars, V64 and V70,
which exhibit eclipses 4-5\% in depth (Fig.~\ref{lc:tr3}). The light
curves of V64 and V70 are consistent with 2.4 and 2.3 $R_J$ objects,
respectively, assuming an orbital inclination of 90$^\circ$. These
stars are located above the cluster MS, especially V70 (on
chip 4), so they are either blends or field stars. 

There are some discrepant points at phase 0.65 in the light curve for
V70. They do not seem to be caused by a defect on the CCD.  It is
possible that we have not recovered the correct period for this
variable. More data would be necessary to resolve this issue.

\begin{figure*}[t]
\epsscale{2.3}
\plottwo{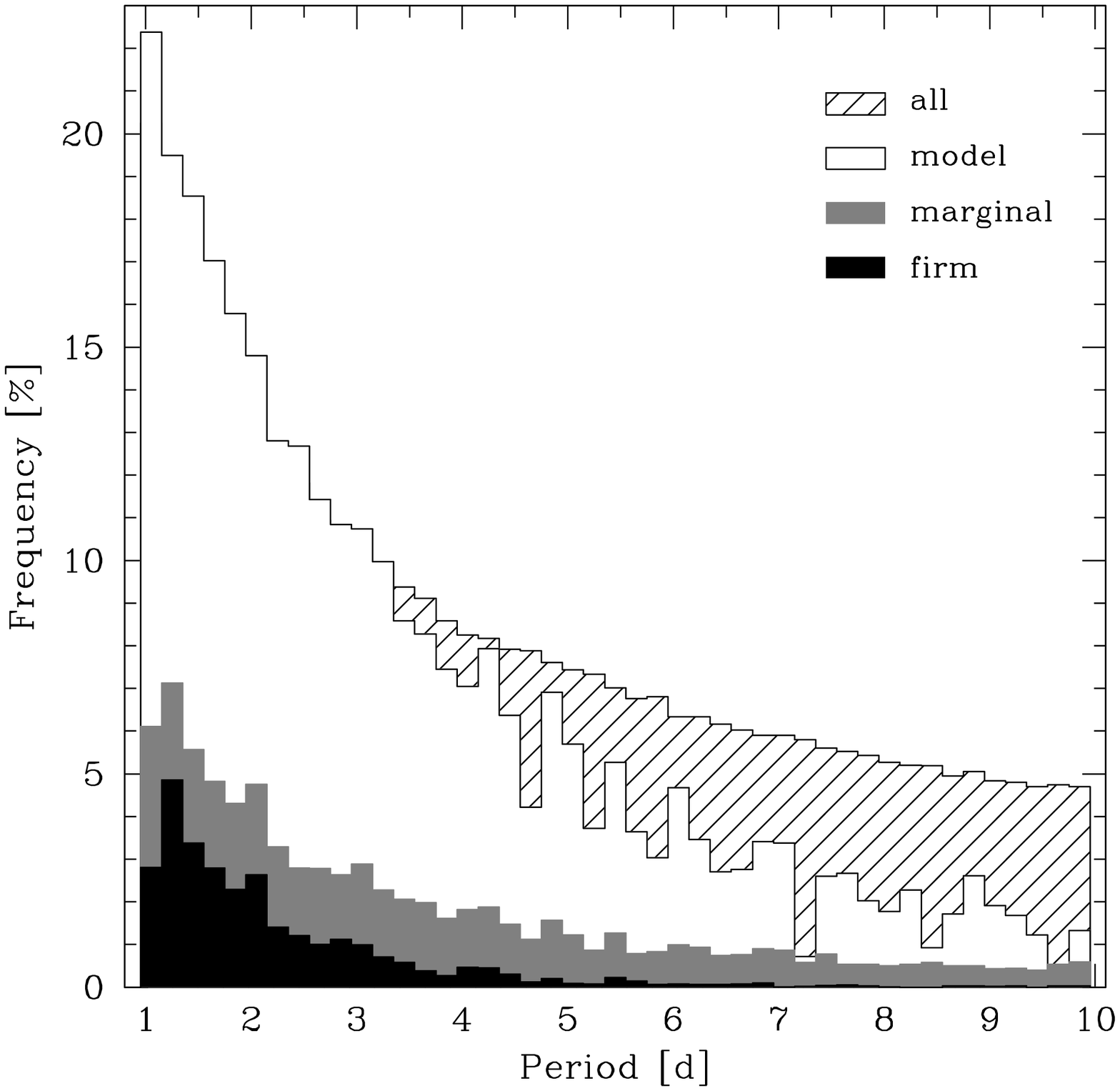}{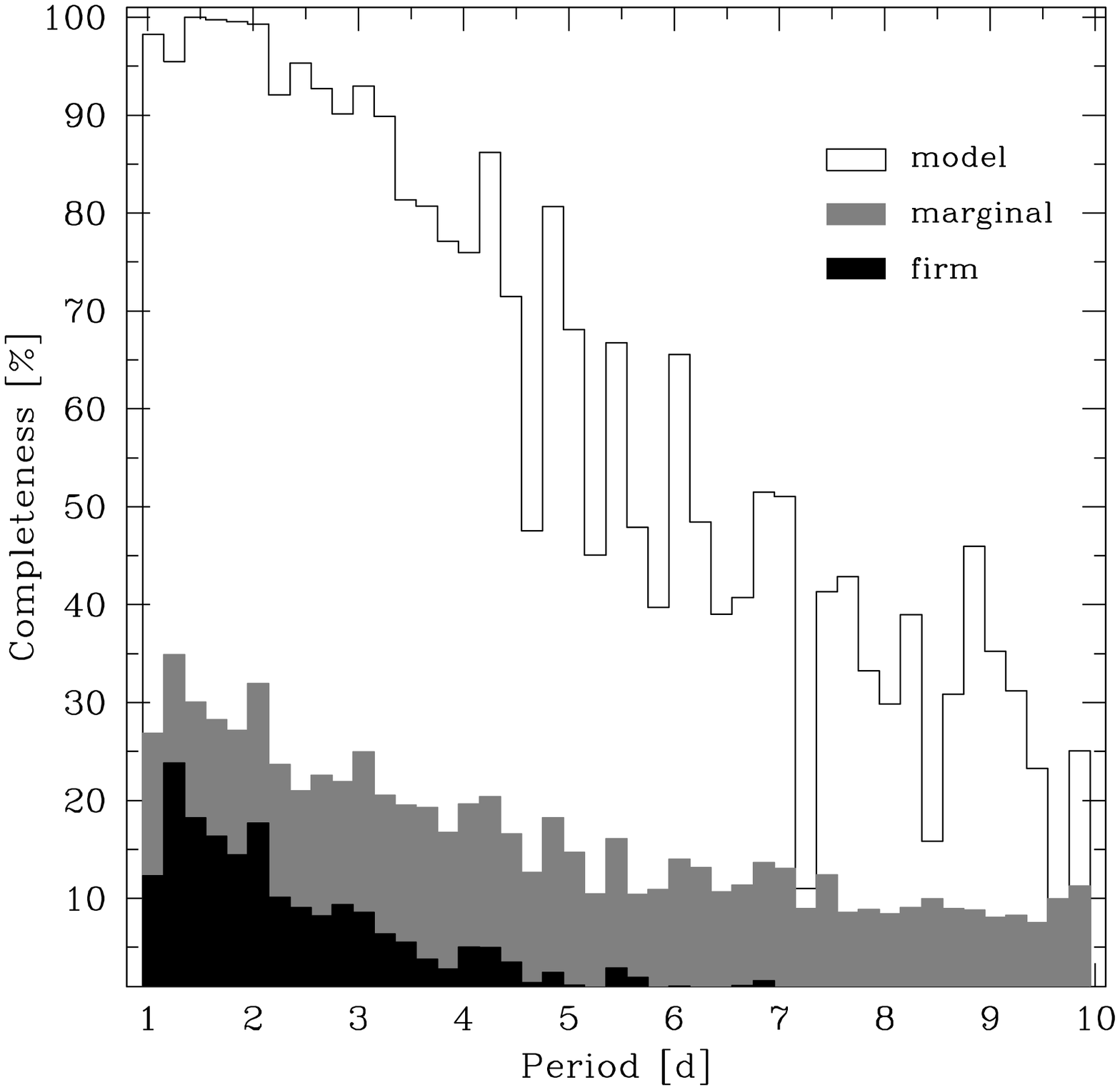}
\caption{Detection efficiency of transiting planets as a function of
their period, relative to planets with all inclinations ($left$) and
all transiting planets ($right$). Shown are the distributions for all
transiting planets ({\it hatched histogram}), detections in the model
light curves ({\it open histogram}) and marginal ({\it filled gray
histogram}) and firm ({\it filled black histogram}) detections in the
combined light curves.}
\label{fig:comp}
\end{figure*}

\section{{\sc Estimate of the Number of Expected Detections}}

The number of transiting planets we should expect to find, $N_P$, can
be derived from the following equation:
\begin{equation}
N_P = N_* f_P D
\end{equation}
where $N_*$ is the number of stars with sufficient photometric
precision, $f_P$ is the frequency of planets within the investigated
period range and $D$ is the detection efficiency, which accounts for
random inclinations. ($N_*$). In \S\S \ref{sec:freq} and
\ref{sec:detef} we determine $f_P$ and $D$.

\subsection{{\it Planet Frequency}}
\label{sec:freq}

The frequency of planets is known to increase with the host star's
metallicity. From Figure~7 in Santos et al.\ (2004), the frequency of
planets with metallicities below [Fe/H] = +0.1 dex is $\sim$2.5\%.
The percentage of planets with periods below 10 days is $15\%$, as
determined in Paper III. Combining these two numbers yields
$f_P$=0.375\%. Please note that the latter estimate is considerably
lower than the commonly adopted frequency of 1\%.

\begin{figure*}[htb]
\plottwo{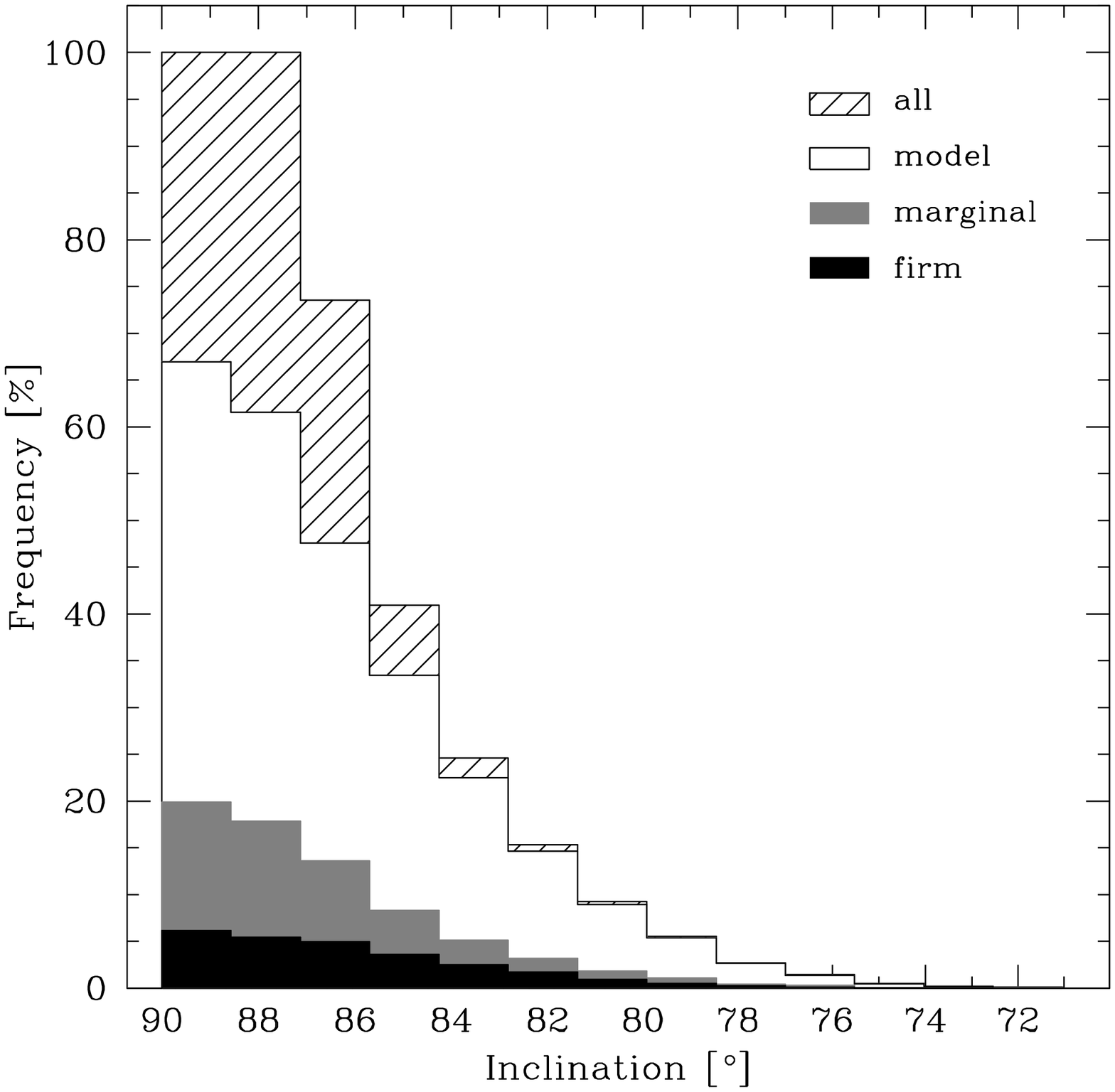}{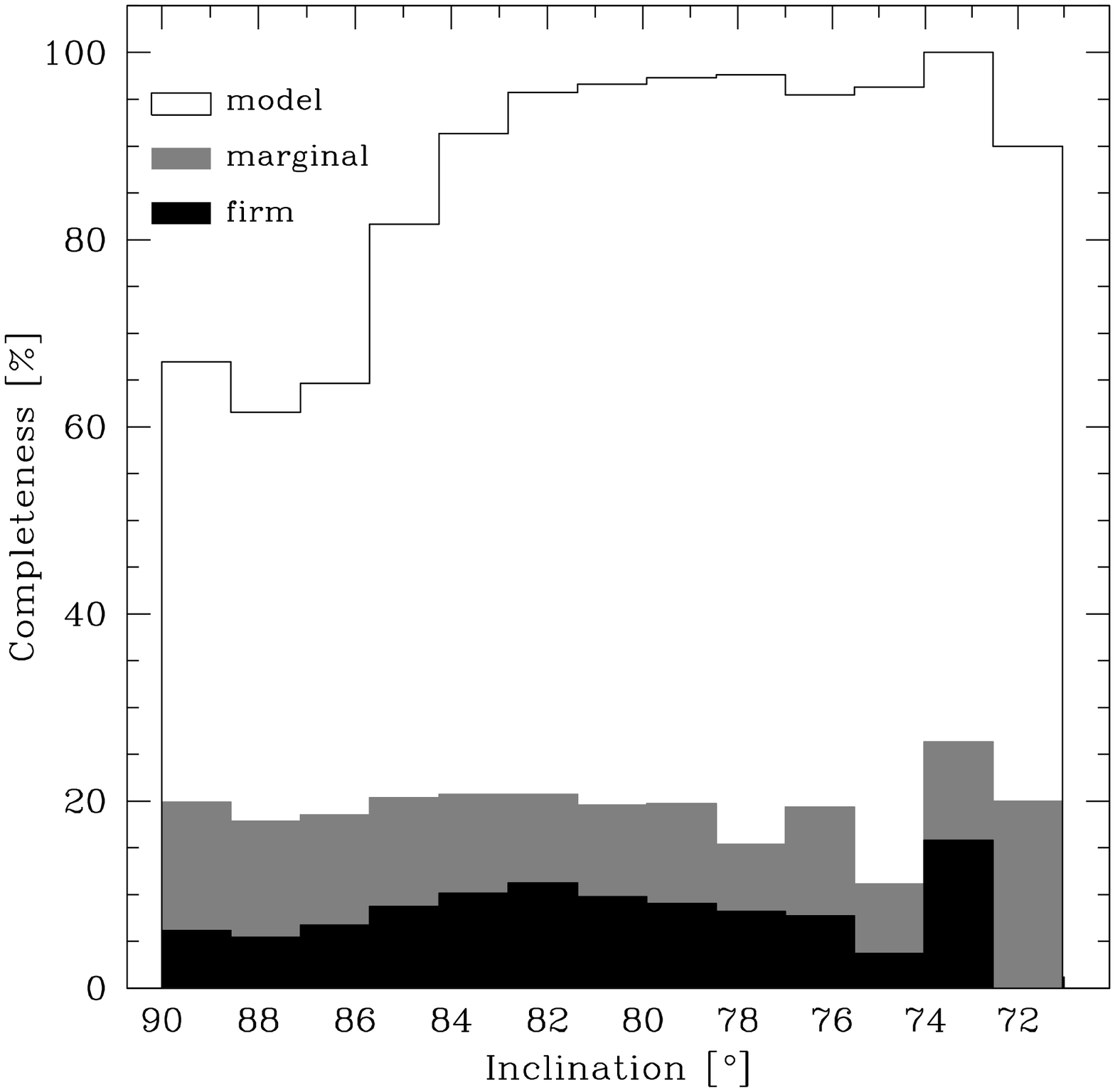}
\caption{Detection efficiency of planetary transits as a function of
their inclination, relative to planets with all inclinations ($left$)
and all transiting planets ($right$).}
\label{fig:i}
\end{figure*}

\begin{figure*}[htb]
\plottwo{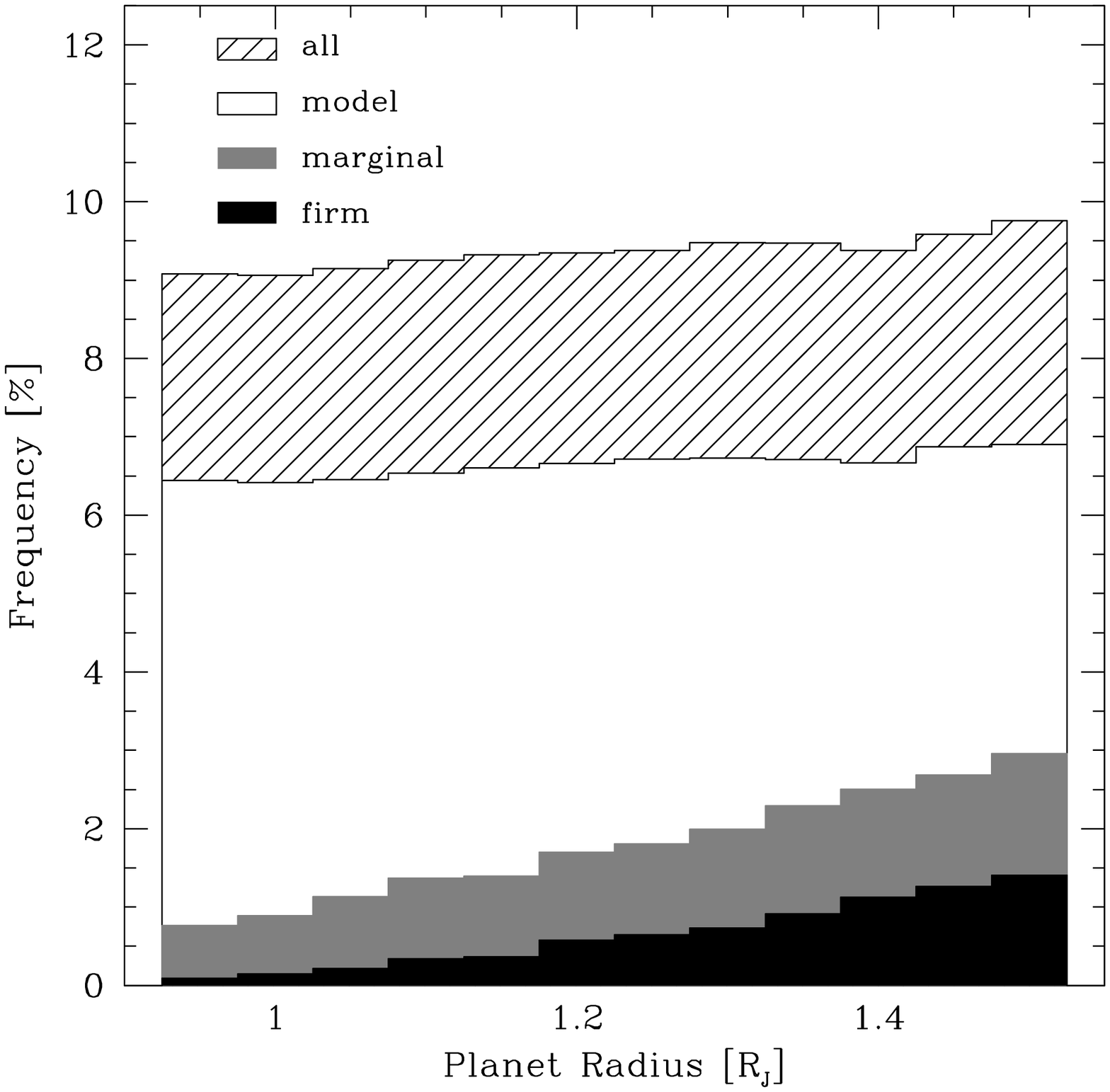}{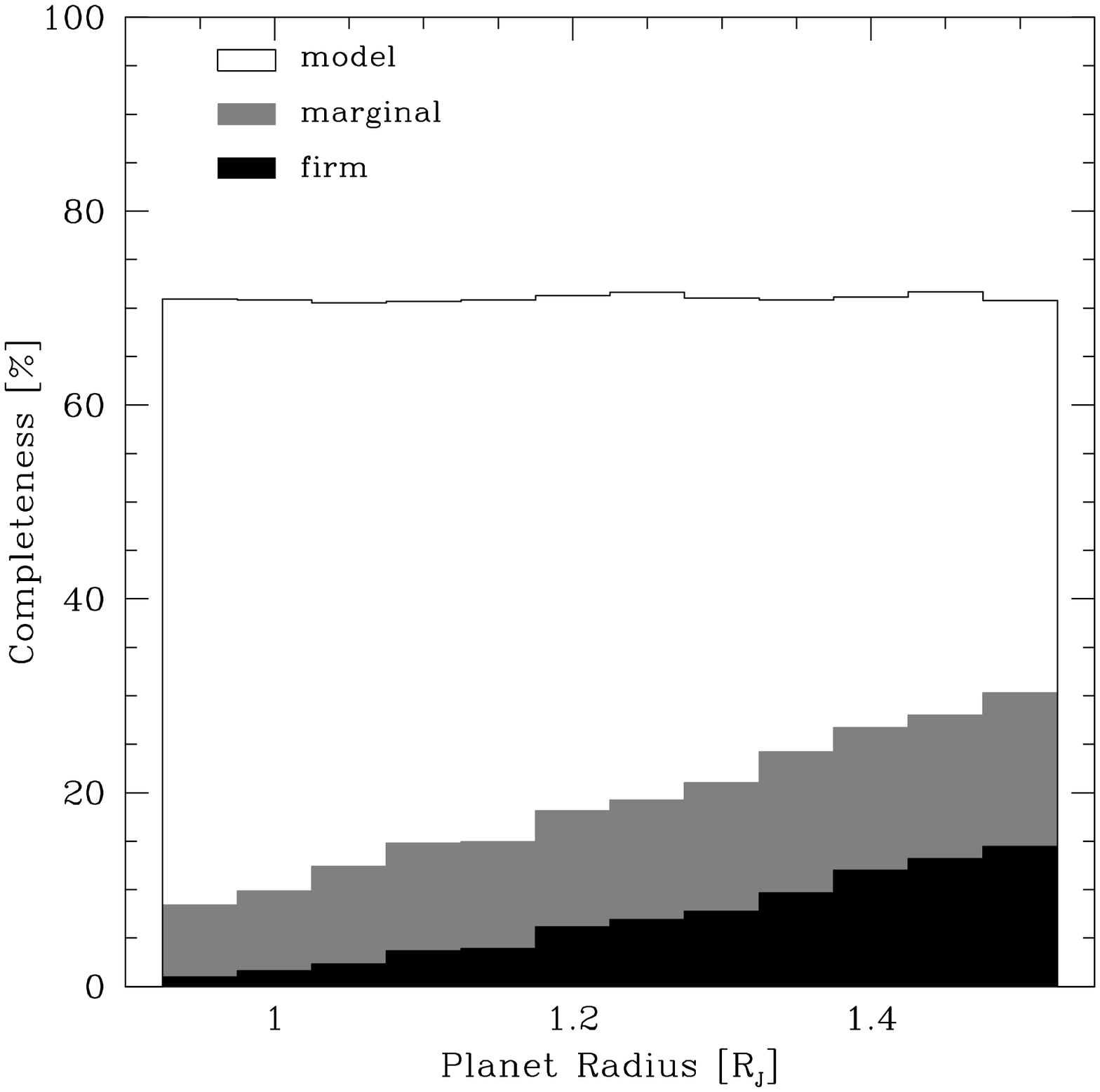}
\caption{Detection efficiency of planetary transits as a function of
their radius, relative to planets with all inclinations ($left$)
and all transiting planets ($right$).}
\label{fig:rp}
\end{figure*}

\begin{figure*}[htb]
\plottwo{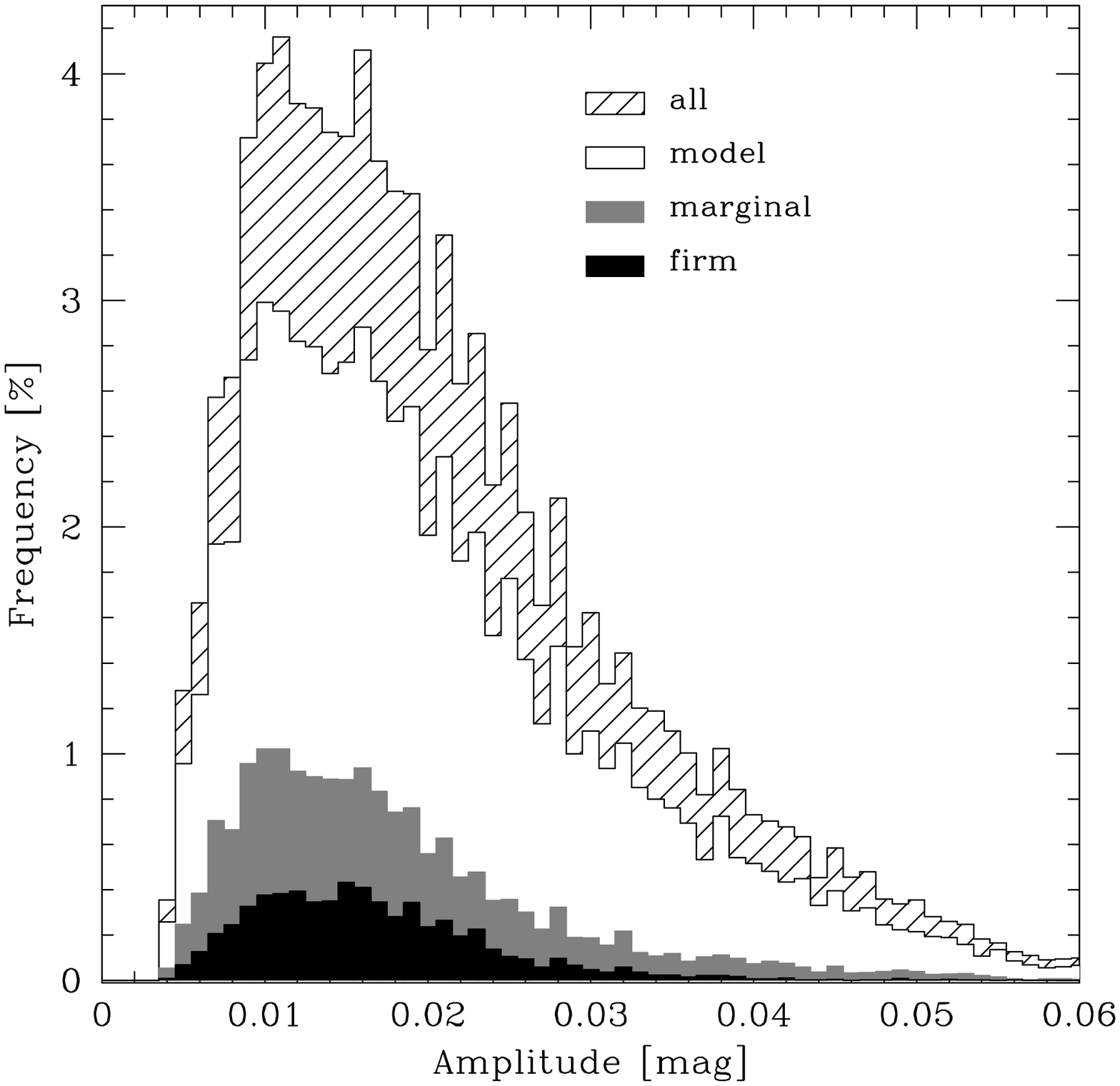}{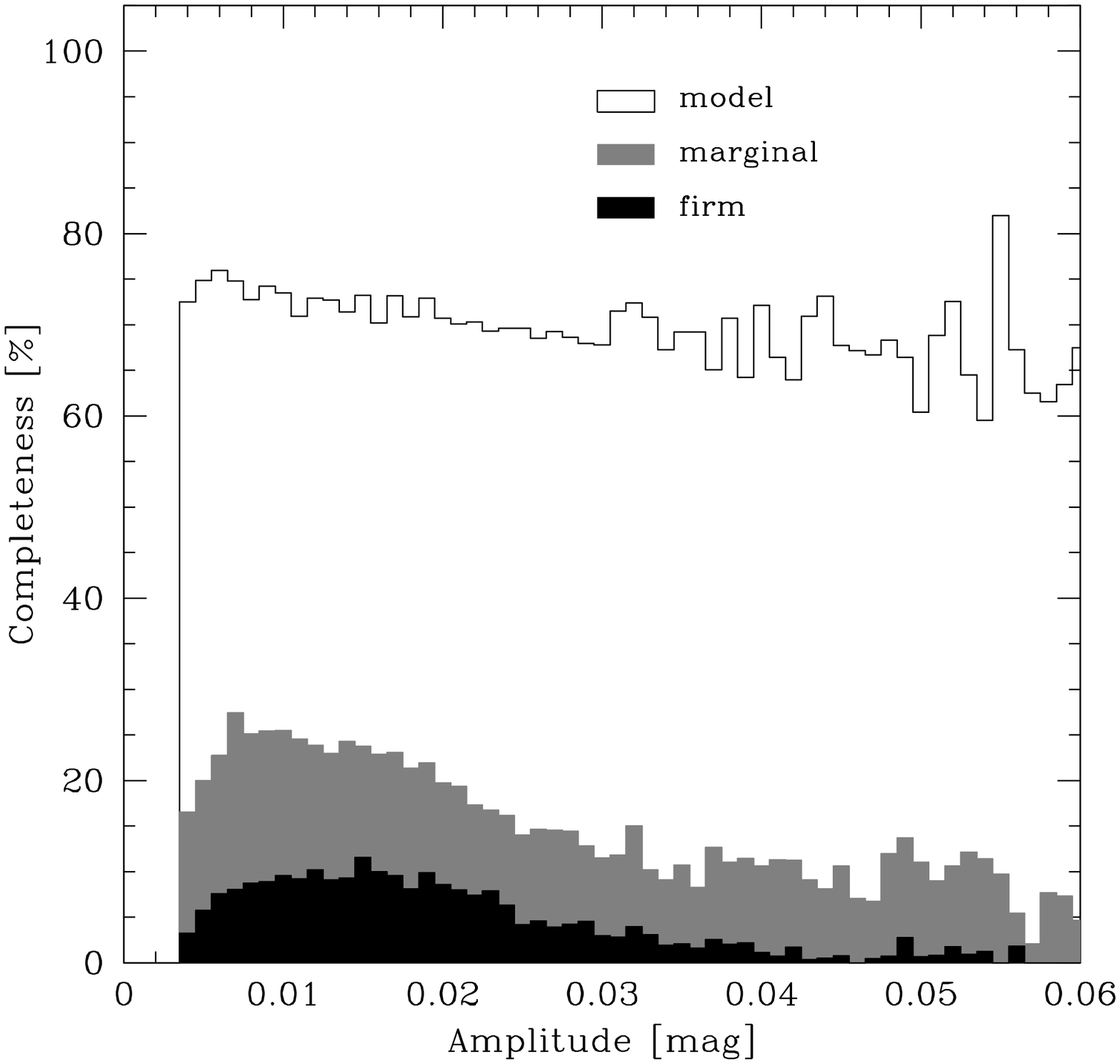}
\epsscale{1.0}
\caption{Detection efficiency of planetary transits as a function of
their amplitude, relative to planets with all inclinations ($left$)
and all transiting planets ($right$).}
\label{fig:amp}
\end{figure*}

\subsection{{\it Detection Efficiency}}
\label{sec:detef}
In order to characterize our detection efficiency, we inserted model
transits into the observed light curves, and tried to recover them
using the BLS method.

\subsubsection{{\it Model Transit Light Curves}}
\label{sec:mod}

The model transit light curves were defined by five parameters: the
transit depth, $\Delta$F, total transit duration, $t_T$, transit
duration between ingress and egress, $t_F$ (the ``flat'' part of the
transit), the period of the planet, $P$ and the limb darkening
coefficient, $u$.

The transit light curves were generated using the approach described
in Paper III. To obtain the radius and mass of the cluster stars, we
used the Z=0.004 1.995 Gyr isochrone of Girardi et al.\ (2000), which
is closest to the Z=0.0048 2 Gyr isochrone used by Ca02. A distance
modulus $(m-M)_R=14.31$ mag was used to bring the observed $R$-band
magnitudes to the absolute magnitude scale (Ca02). 

In addition to $P$, the equations contain two other free parameters:
the planet radius, $R_P$ and the inclination of the orbit, $i$. A
fourth parameter which affects the detectability of a planet is the
epoch of the transits, $T_0$.

\subsection{\it Test Procedure}

We investigated the range of parameters specified in Table
\ref{tab:pars}, where $P$ is expressed in days, $R_P$ in $R_J$, $T_0$
as a fraction of period. We examined the range of periods from 1.05 to
9.85 days and planet radii from 0.95 to 1.5 $R_J$, with a resolution of
$0.2$ days and $0.05$ $R_J$, respectively. For $T_0$ we used an
increment of 5\% of the period, and a 0.025 increment in $\cos i$. The
total number of combinations is 432000.

We followed the test procedure described in Paper III. The tests were
run on the 5476 stars from chips 1-4 with at least 500 good epochs and
light curve rms below 0.05 mag. Figure~\ref{fig:tr} shows the
original, model and combined light curves (upper, middle and lower
panels, respectively) for a star with $R=17.21$ and a planet with a
period of 2.05 days, radius of 1.45 $R_J$ and inclination of
84$^{\circ}$. The amplitude of the transit is 0.0175 mag, and the mass
and radius of the star, taken from the models, are 1.18 $M_{\odot}$
and 1.10 $R_{\odot}$.

To assess the impact of the procedure to correct for offsets between
the runs on our detection efficiency, we investigated three cases,
where the correction was applied:
\begin{itemize}
\vspace{-0.2cm}
\item [A.]{after inserting transits,}
\vspace{-0.2cm}
\item [B.]{before inserting transits,}
\vspace{-0.2cm}
\item [C.]{was not applied at all.}
\vspace{-0.2cm}
\end{itemize}
Case (B) will give us the detection efficiency if our data did not
need to be corrected, and case (C) if we did not apply the
corrections. Case (A) will give us our actual detection efficiency,
and its comparison with cases (B) and (C) will show how it is affected
by the applied correction procedure.

\subsection{\it Detection Criteria}

The same detection criteria as in Paper III were used. A transit was
flagged as detected if:
\begin{enumerate}
\vspace{-0.2cm}
\item {The period recovered by BLS was within 2\% of the input period
$P_{inp}$, 2 $P_{inp}$ or $\frac{1}{2}$ $P_{inp}$,}
\vspace{-0.2cm}
\item {The BLS statistics were above the following thresholds: $SDE >
6$, $\alpha > 9$.}
\vspace{-0.2cm}
\end{enumerate}
These detections will be referred to hereafter as {\it firm}.
Detections where only condition (1) was fulfilled will be called {\it
marginal}.

\subsection{\it Detection Efficiency}
\label{subsec:detef}
The results of the tests are summarized in Table~\ref{tab:art}, which
lists the test type (A-C), the number and percentage of transits with
$t_T \geq 0.5^h$ (out of the 432000 possible parameter combinations),
and the numbers and percentages (relative to the total number of
transits in column 2) of transits detected in the model light curves,
and of marginal and firm detections in the combined light curves.

Figures~\ref{fig:comp}-\ref{fig:amp} show the dependence of the
detection efficiency on period, inclination, planet radius and transit
amplitude. The hatched, open, filled gray and filled black histograms
denote distributions for all transiting planets, planets detected in the
model light curves, and marginal and firm detections in the combined
light curves, respectively. Left panels show the frequency of transits
and transit detections relative to planets with all inclinations. Right
panels show the detection completeness normalized to all transiting
planets (plotted as hatched histograms in left panels).

The tests show that 9\% of planets with periods 1-10 days will
transit their parent stars. This frequency drops from $\sim$22\% at
$P=1^d$ to $\sim$5\% at $P=10^d$. All planets with inclinations
$87-90^\circ$ transit their host stars, and this fraction drops to
$\sim$78\% for $i=86^\circ$ and $\sim$5\% for $i=79^\circ$. The
frequency of transits increases very weakly with planet radius. The
distribution of transit amplitudes has a wide peak stretching from
0.8\% to 2\%, centered on $\sim$1.4\%.

The percentage of detections for the model light curves illustrates
the limitation imposed on our detection efficiency by the temporal
coverage alone. Due to incomplete time sampling, we are restricted to
71\% of all planets with periods between 1 and 10 days. For periods
below 3 days, our temporal coverage is sufficient to detect $\sim$90\%
of all transiting planets, and drops to $\sim$50\% at $P=7$ days. The
detection completeness increases with decreasing inclination because
at lower $i$ only short period planets can transit their host
stars. It does not depend on the planet radius, and it decreases with
increasing transit amplitude.

The source of the dependence of the detection completeness on transit
amplitude is not as straightforward as for the other correlations. The
amplitude depends on the radii of the star and planet. Since the
detection completeness was found to be largely independent of the
planet radius, the observed trend must stem from its dependence on the
host star's radius, which is a function of its magnitude. Such a
correlation is indeed observed, with completeness increasing for
brighter stars (not shown here). The link between the temporal
coverage and magnitude comes from the observed increase in the number
of points in the light curve with decreasing magnitude.

For cases A, B and C, we {\it marginally} detect 19\%, 19\% and 12\%
of all transiting planets, and {\it firmly} detect 7.0\%, 6.6\% and
2.6\%, respectively. Transiting planets with firm detections
constitute 80\%, 82\% and 80\% of all stars with $SDE > 6$ and $\alpha
> 9$. Correcting the light curves after inserting transits (case A)
produces almost the same number of detections, compared to the case
where the correction was applied to the original light curves (case
B).  If the light curves were not corrected (case C), we would detect
only 37\% of the transiting planets detected in case A.

The detection completeness for firm detections peaks at 20\% for
periods $1-2^d$ and decreases with period more steeply than model
detections. It does not show a marked dependence on inclination, and
strongly increases with increasing planet radius, from $\sim 2\%$ at 1
$R_J$ to 15\% at 1.5 $R_J$. This is also apparent in the right
panel of Fig.~\ref{fig:rms}, which shows the detection efficiency of
0.95, 1.15, 1.30 and 1.50 $R_J$ planets as a function of magnitude
and rms scatter (white: 0\%, black 100\%).

The detection efficiency peaks at an amplitude of $\sim$1.5\%, due to
the most favorable ratio between the transit amplitude and photometric
accuracy for this amplitude/magnitude range.

The efficiency of {\it firm} transiting planet detections, relative to
planets with all orbital inclinations, $D$, is $2818/432000=0.65\%$.

\begin{figure*}[!htb]
\epsscale{2.0}
\plotone{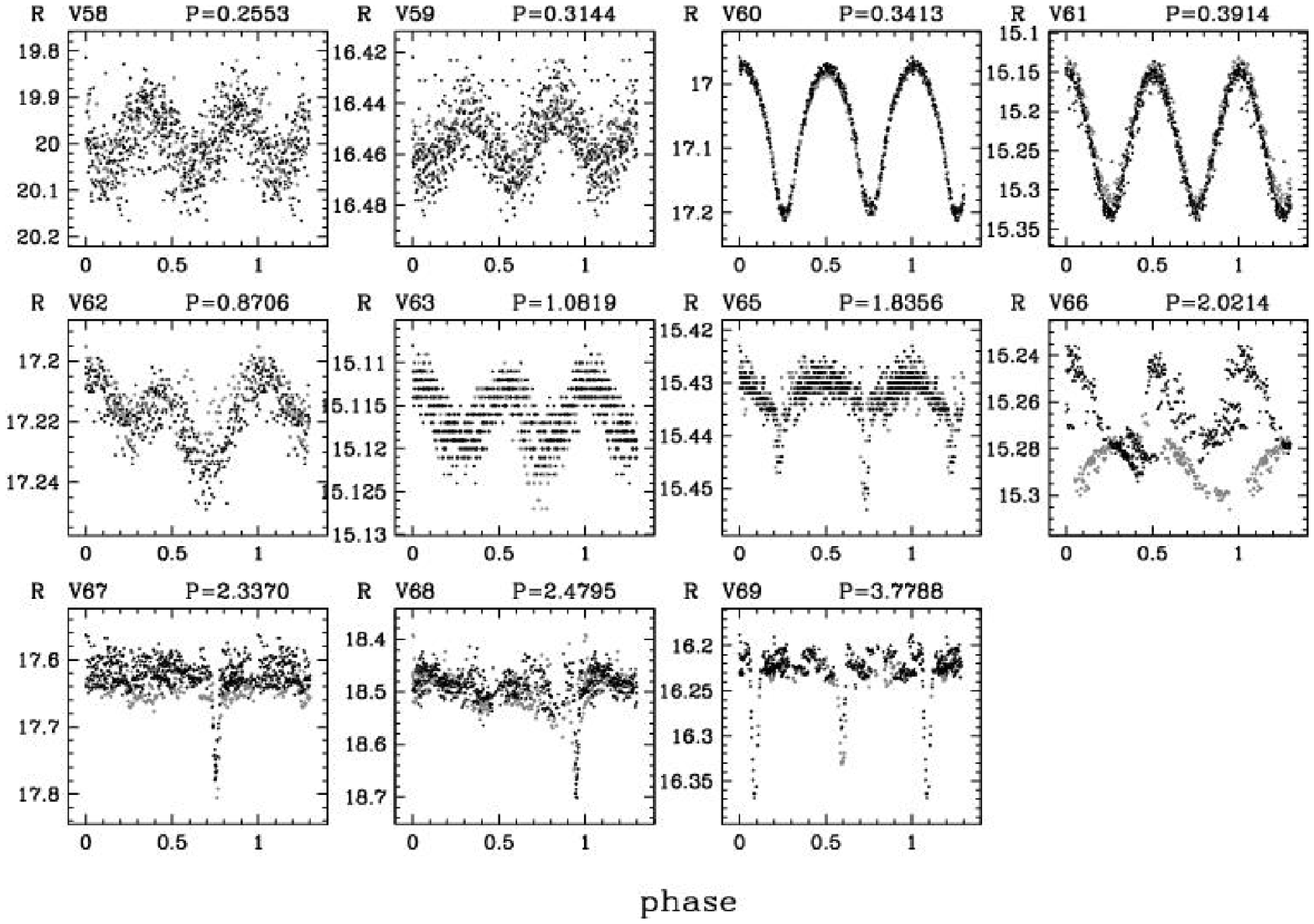}
\caption{$R$-band light curves of 11 new eclipsing binaries. Data
points from the first observing season are denoted by gray symbols
and from the second season by black symbols.}
\label{lc:ecl}
\end{figure*}

\subsection{\it Number of Transiting Planets Expected}

In \S\S~\ref{sec:freq}-\ref{sec:detef} we determined the planet
frequency $f_P$ to be $0.375\%$, the number of stars as 5159 and and
our detection efficiency $D$ as $0.65\%$. We should thus expect 0.13
transiting planets among the cluster and field stars.

\subsection{\it Discussion}

Figure~\ref{fig:comp} demonstrates that our temporal coverage is not the
limiting factor. To increase the number of expected planets it would be
necessary to improve the photometric precision.  The weather and seeing
conditions turned out to be inferior to what we were expecting. A better
quality CCD and a telescope with a larger diameter and/or better
observing conditions would be required to improve the chances for a
successful transiting planet search in NGC~2158.

The precision of this estimate is largely limited by the uncertainty
in one of our basic assumptions -- the distribution of planetary
radii.  This distribution is not precisely known, and changing it will
have a marked effect on the final result. Adopting a distribution of
planetary radii from 1.0 to 1.35 $R_J$, corresponding to the radius
range spanned by the seven known Jupiter-mass transiting planets (The
Extrasolar Planets Encyclopaedia) would lower D from 0.65\% to
0.33\%. This translates to 0.06 detections, compared to 0.13 with the
original radius distribution -- a 50\% decrease.

In Paper I we made the assumption that the planetary radii would span
the range 1-3 $R_J$, based on the radius of $1.347$ $R_J$ for the only
known transiting planet at the time, HD 209458b (Brown et al.\
2001). A simulation for planets in the radius range 1.5-3.0 $R_J$
shows that 10\% of them transit their parent stars, 72\% are detected
in the model light curves, and 52\% and 35\% are marginal and firm
detections in the combined light curves. Assuming that planet radii
are distributed evenly between 1 and 3 $R_J$ would give the percentage
of firm detections of 28\% and detection efficiency $D=2.9\%$, which
translates into 0.56 expected detections. 

We have detected a star which, if a member of the cluster and not a
grazing binary, has a transiting companion with a likely radius of
about 1.7 $R_J$. The detection rate for 1.7 $R_J$ objects is 21\%.
This would imply that $0.09$\% of stars in our field of view have a
transiting companion of this mass at close separation (period below 10
days). 

\section{\sc Variable Stars}
We extracted the light curves of 57 known variable stars, discovered
in Paper II. In Tables \ref{tab:ecl}-\ref{tab:misc} we list their
revised parameters.

\begin{figure*}[!ht]
\plotone{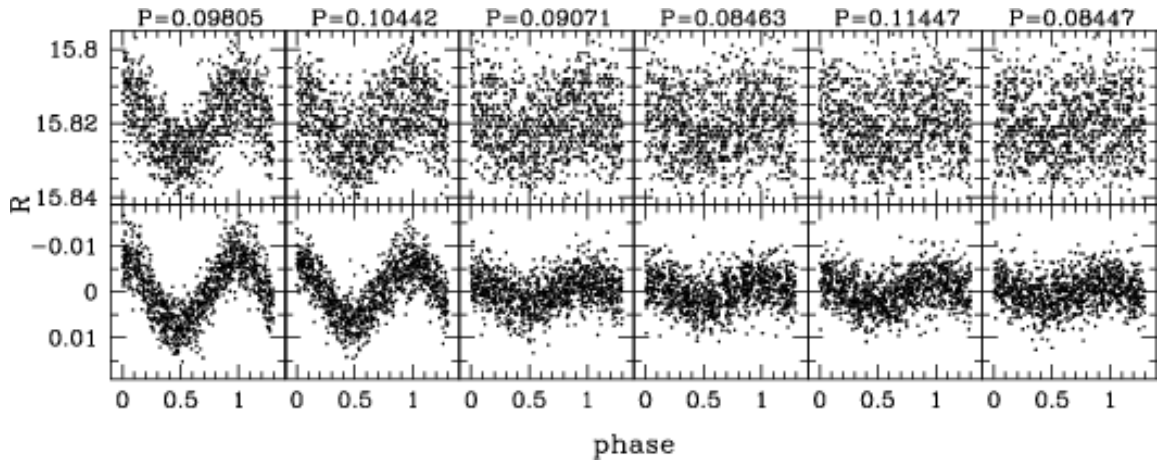}
\caption{$R$-band light curves of $\delta$ Scuti variable V35, phased
with each of the six detected periods: $top$, raw; $bottom$,
subtracted of variability corresponding to other periods.}
\label{lc:v35}
\end{figure*}

\begin{figure*}[!ht]
\plotone{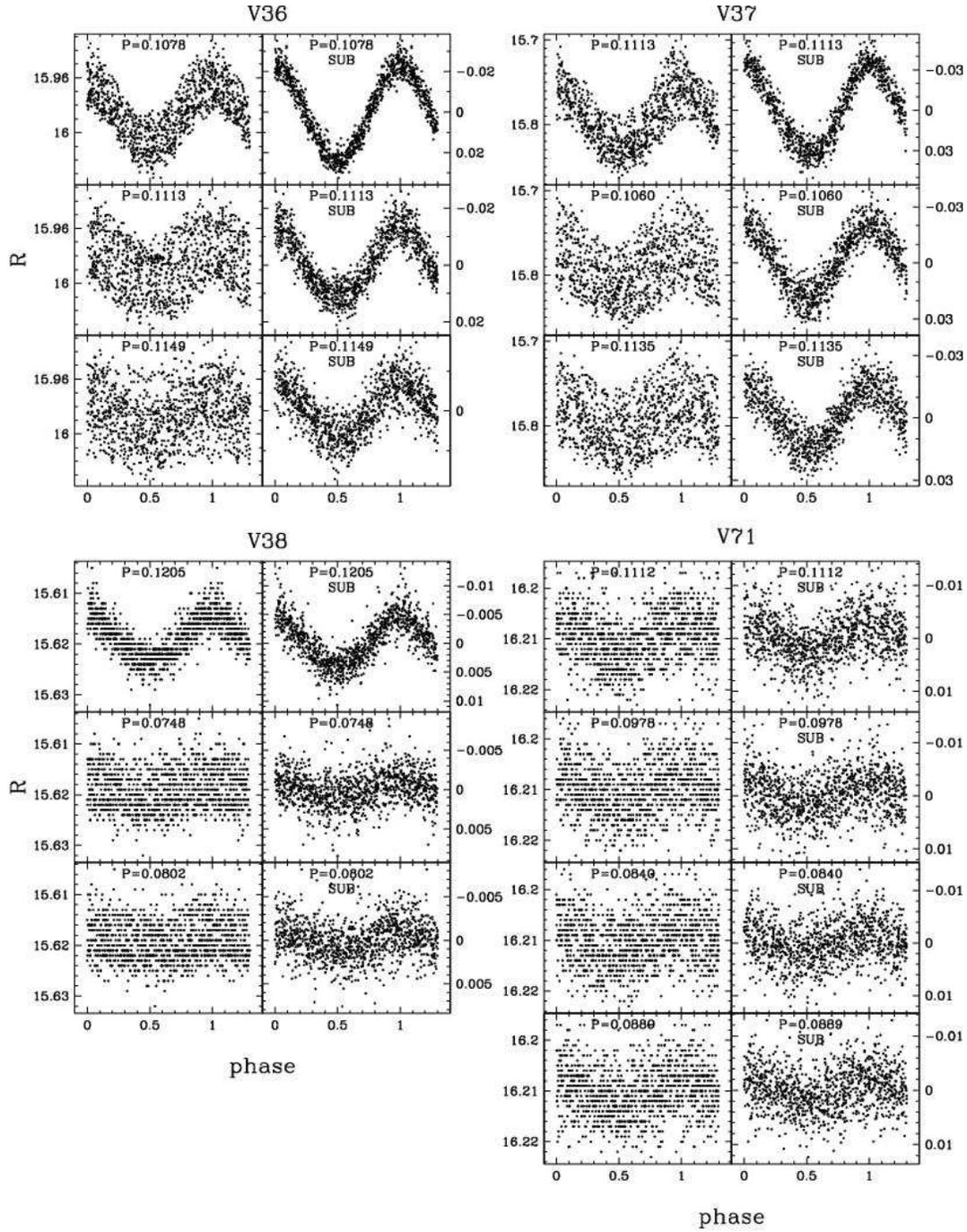}
\caption{$R$-band light curves of $\delta$ Scuti variables V36, V37,
V38 and V71, phased with each detected period: $left$, raw;
$right$, subtracted of variability corresponding to other periods.}
\label{lc:dsct}
\end{figure*}

\begin{figure*}[hp]
\plotone{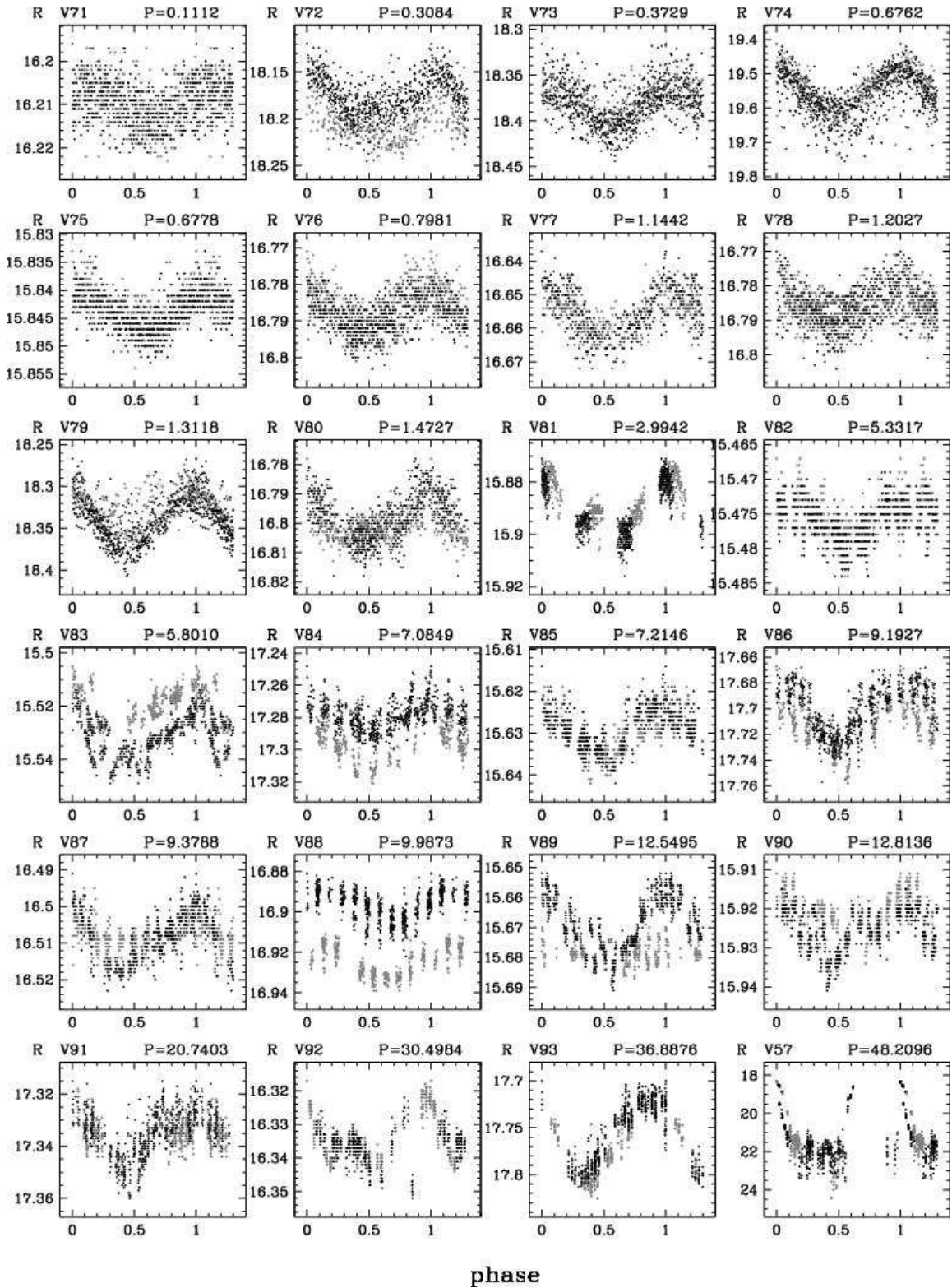}
\caption{$R$-band light curves of 23 new periodic variables and the
light curve of the cataclysmic variable V57 phased with the proposed
cycle length of 48.2 days. Data points from the first observing season
are denoted by gray symbols and from the second season by black
symbols.}
\label{lc:per}
\end{figure*}

\begin{figure*}[!ht]
\plotone{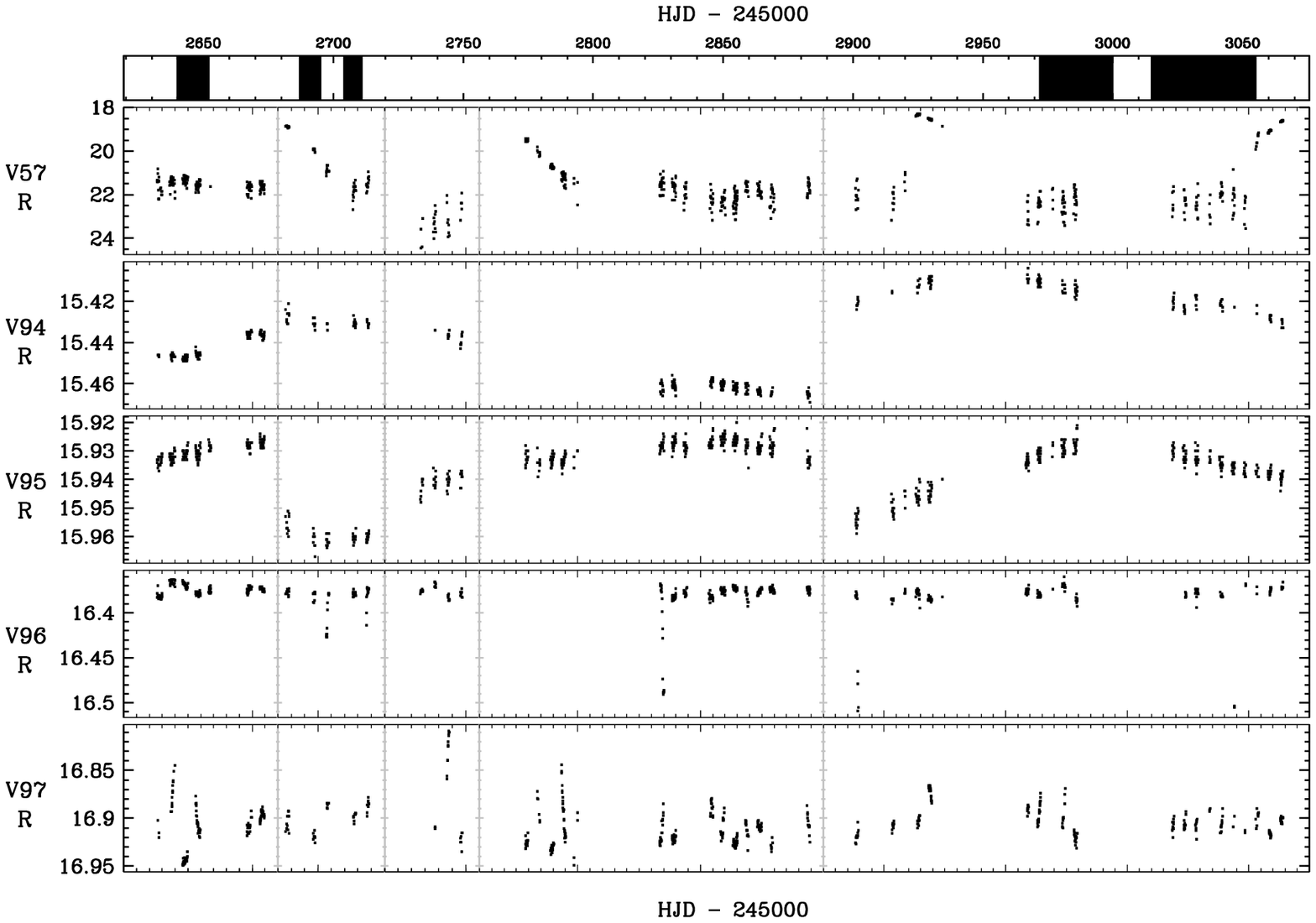}
\caption{$R$-band light curves of miscellaneous variables. The top
window illustrates the distribution in time of the four sub-windows
plotted for the variables.}
\label{lc:misc}
\end{figure*}

We searched for new variables by running BLS in the period range
0.1-10 days. We have discovered 40 new variables: 2, 6, 24 and 8 on
chips~1-4, respectively: 13 eclipsing binaries, 23 other periodic
variables and four long period or non-periodic variables.  Their
parameters are listed in Tables~\ref{tab:ecl}-\ref{tab:misc} and their
light curves are shown in Figures~\ref{lc:tr3} and
\ref{lc:ecl}-\ref{lc:misc}.\footnote{The $VR$ band photometry and
finding charts for all variables are available from the authors via
the anonymous ftp on cfa-ftp.harvard.edu, in the /pub/bmochejs/PISCES
directory.}  They are also plotted on the CMD in Fig.~\ref{fig:cmd}.

\subsection{Eclipsing Binaries}

There are 13 eclipsing binaries among the newly discovered variables
(Fig.~\ref{lc:tr3} and \ref{lc:ecl}, Tab.~\ref{tab:ecl}).  V58-V63 are
W UMa type binaries. V60 is very likely a cluster member, based on the
KAS97 membership probability and its location on the binary MS. V61
and V63 are located in the blue straggler region. Variables V64 and
V67-V70 appear to be detached systems. V64 and V70 were discussed
previously in \S \ref{sec:cand}. V65 displays continuous variation
between the eclipses. It appears to be a cluster member and is located
at the base of the red giant branch. The shape of the light curve of
V66 changed dramatically between 2003 and 2004. This star is a blue
straggler, if it belongs to the cluster.

\subsection{$\delta$ Scuti Stars}

In Paper II we reported the discovery of five $\delta$ Scuti
variables, V35-V39. V71 is a newly discovered variable of this
type. We have identified four periods in its light curve. It is
located just above the MSTO, together with the other
$\delta$ Scuti variables (Fig.~\ref{fig:cmd}).

Table~\ref{tab:pul} lists the parameters and Table~\ref{tab:mm} the
periods derived for the $\delta$ Scuti variables. Figures~\ref{lc:v35}
and \ref{lc:dsct} show their light curves phased with every identified
period. We have identified at least three modes of pulsation for all
variables except V39. The identification of the mode of pulsation in
$\delta$ Scuti stars is not trivial, as the frequencies of highest
amplitude have very different order in different stars (Arentoft et
al.\ 2005). The last two columns of Table \ref{tab:mm} give the
minimum and maximum difference between the distance modulus computed
from the fundamental-mode period-luminosity relation for $\delta$
Scuti stars (eq. [4] in Petersen \& Christensen-Daalsgard 1999) and
the distance modulus (m-M)$_0$=12.80 derived by Ca02. These
differences are consistent with the variables being members of the
cluster, with the exception of V39. It is likely that the discrepancy
for this variable is a result of non-fundamental mode pulsations.

As discussed in Paper II, four of these stars have proper motion data
in KAS97. Cluster membership of V36 is confirmed with high
probability, while for V35 and V39 the data are not conclusive. The
catalog reports a very low cluster membership probability for V38,
which is in disagreement with our conclusions based on the $\delta$
Scuti P-L relation. The study of KAS97 was also targeting much brighter
stars ($V>8$ mag) in M~35 and two other clusters at smaller distances
than NGC~2158. The authors say that their proper motion measurement
``accuracy decreases rapidly for stars fainter than $V=15.5$ mag'',
while the $V$ magnitude range of NGC~2158 variables is 15.7-18.0, and
$V=16.04$ for V38.

\subsection{Other Periodic Variables}

We have discovered 22 new other periodic variables (Fig.~\ref{lc:per}
and Tab.~\ref{tab:pul}). They display roughly sinusoidal light
variations with periods ranging from 0.3 to 36 days. The variability
in most of these stars is probably caused by star spots, which are
rotating in and out of view. Variables V83, V84, V88 and V89 in
addition to periodic variability show a change in the magnitude
zeropoint between the 2003 and 2004 observations, which is most likely
caused by the evolution of star spots.

On the CMD, V75 and V81 are located in the blue straggler region of
the CMD, V76, V77, V78, V80 and V87 are located close to MSTO and V82,
V83, V85 and V89 are located near the sub-giant branch (SGB).  The
variable V90, if it belongs to the cluster, might be a member of the
newly proposed class of variable stars termed ``red stragglers''
(Albrow et al.\ 2001) or ``sub-subgiant stars'' (Mathieu et al.\
2003). Thus far, the origin and evolutionary status of these stars
remains unknown. V79 and V91 are located above the MS and may be
binary stars or background objects and V72, V73, V74, V84, V86, V88
and V92 seem to be background objects. According to KAS97, V76 is most
likely a cluster member, and for V75, V81 and V88 the data are not
conclusive.

\subsection{Cataclysmic Variable Candidate}

In Figure \ref{lc:misc} we present the light curve of the cataclysmic
variable candidate, V57, identified in Paper II. The variable shows
four 2.5 mag outbursts which phase with a 48.2 day period (phased
light curve shown in Fig.~\ref{lc:per}).

In Paper II we determined the $V-R$ color of this variable to be -0.26
in maximum and -0.31 in minimum. In our current data the variable has a
much more plausible $V-R$ color: 0.304 in maximum and 0.874 in minimum.
The color determination at minimum is very uncertain, due to the
faintness of the star and differing minimum magnitudes between the
cycles. 

Assuming the NGC~2158 distance and reddening derived by Ca02, we get
the minimum absolute $V$-band magnitude $M_V(min)$ of 6.4 mag. From
Fig.~3.5 in Warner (1995) it is apparent that $M_V(min)$ is roughly 8
mag for U Gem type dwarf novae, but even the most extreme Z Cam type
systems have $M_V(min)>6.8$ mag. This would imply that V57 is a
dwarf nova located in front of the cluster.

It should be noted, however, that the $V$ band magnitude alone is not
sufficient to exclude this variable as a cluster member. Another
cataclysmic variable, B7 in the open cluster NGC~6791, is seen most of
the time in a high state at $M_V\sim5$, and has been caught only once
in its low state at $M_V(min)=7.4$ (Kaluzny et al.\ 1997). The high
state in VY Scl type CVs can last for several years, and some systems
stay in the high state during most of the time (i.e. ST LMi, MR Ser
and AN UMa in Kafka \& Honeycutt\ 2005). Further investigation is
needed to resolve the nature of V57. A spectrum of a dwarf nova in
quiescence should display Balmer emission on a blue continuum. During
outburst the emission lines are gradually overwhelmed by the
increasing continuum and development of broad absorption lines (Warner
1995). If V57 is in a ``high'' level, like B7, then this will be
apparent in the spectrum and emission lines should not be prominent.

\subsection{Long Period and Non-Periodic Variables}

We have discovered four new long period or non-periodic variables
(Fig.~\ref{lc:misc}, Tab.~\ref{tab:misc}). V94
is located near the SGB branch.  V95 is located in the red starggler
region of the CMD. According to KAS97, V95 is likely a cluster member.
V96 and V97 are located on the upper MS, close to the MSTO. V96 is a DEB
and could be useful for cluster distance and age determination, due to
its location on the MSTO.

\section{{\sc Conclusions}}
In this paper we have performed an extensive search for transiting
planets in the intermediate age, populous cluster NGC~2158. The
cluster was monitored for over 260 hours during 59 nights. 

We have identified a low-luminosity transiting object candidate, TR1.
The $R$-band amplitude of 3.7\% implies a 1.66 $R_J$ radius for the
transiting companion. The location of the host star on the cluster
MS and its proximity to the cluster center seem to indicate
that it is a member of the cluster. Higher accuracy light curves are
required to better costrain the radius and period of TR1, and followup
spectroscopy, to estimate the mass of the transiting object, through a
measurement, or an upper limit on the central object's radial velocity
variations.

Assuming a planet frequency from radial velocity surveys, we estimate
that we should have detected 0.13 transiting planets with periods
between 1 and 10 days, with our photometric precision and temporal
coverage. The main limitation on our detection efficiency was imposed
by the photometric precision.

We have discovered 40 new variable stars in NGC 2158: 13 eclipsing
binaries, 23 other periodic variables and four non-periodic variables,
Together with 57 variables discovered in Paper II, this brings the
total number of variables known in this cluster to 97. We have also
presented high photometric precision light curves, spanning almost 13
months, for all previously known variables.

Transiting planets have proven to be more challenging to detect than
initially expected, as shown by the paucity of detections from the
many searches under way in open clusters (i.e.\ Bruntt et al.\ 2003;
UStAPS: Hood et al.\ 2005; EXPLORE/OC: von Braun et al.\ 2004; STEPSS:
Marshall et al.\ 2005) and in the Galactic field (i.e.~EXPLORE:
Mall{\' e}n-Ornelas et al.\ 2003; OGLE: Udalski et al.\ 2002a; STARE:
Alonso et al.\ 2003; HAT: Bakos et al.\ 2004; WASP0: Kane et al.\
2005\footnote{For a more complete list of transiting planet searches,
please refer to {\tt
http://star-www.st-and.ac.uk/\~{}kdh1/transits/table.html} and {\tt
http://www.obspm.fr/encycl/searches.html}}). To date, only six
planets have been discovered independently by transit searches, all of
them in the field, and five of those were initially identified by OGLE
(Udalski et al.\ 2002a, 2002b, 2002c, 2003; Alonso et al.\ 2004).

\acknowledgments{ We would like to thank the FLWO 1.2 m Time
Allocation Committee for the generous amount of time we were allocated
to this project.

This publication makes use of data products from: the Two Micron All
Sky Survey, which is a joint project of the University of
Massachusetts and the Infrared Processing and Analysis
Center/California Institute of Technology, funded by the National
Aeronautics and Space Administration and the National Science
Foundation; the Digital Sky Survey, produced at the Space Telescope
Science Institute under U.S. Government grant NAG W-2166; the SIMBAD
database, operated at CDS, Strasbourg, France and the WEBDA open
cluster database maintained by J.~C.~Mermilliod
(http://obswww.unige.ch/webda/).

Support for BJM and JNW was provided by NASA through Hubble Fellowship
grants HST-HF-01155.02-A, HST-HF-01180.01-A from the Space Telescope
Science Institute, which is operated by the Association of
Universities for Research in Astronomy, Incorporated, under NASA
contract NAS5-26555. KZS acknowledges support from the William
F.~Milton Fund.}

\begin{deluxetable}{rrrrrrrr}
\tabletypesize{\footnotesize}
\tablewidth{0pc}
\tablecaption{Low Luminosity Transiting Object Candidate in NGC 2158}
\tablehead{\colhead{ID} & \colhead{$\alpha_{2000}$ [h]} &
\colhead{$\delta_{2000}$ [$\circ$]} &\colhead{P [d]} & \colhead{$R_{max}$} &
\colhead{$V_{max}$} &\colhead{$A_R$} & \colhead{$A_V$} }
\startdata
TR1 &  6 07 35.4 & 24 05 40.8 &  2.3629 & 18.544 & 19.218 & 0.037 & \nodata \\
\enddata
\label{tab:tr}
\end{deluxetable}

\begin{deluxetable}{lllcc}
\tabletypesize{\footnotesize}
\tablewidth{0pt}
\tablecaption{Parameter Range}
\tablehead{
\colhead{Parameter} & \colhead{min} & \colhead{max} & \colhead{step}
& \colhead{n$_{steps}$}}
\startdata
P (days)      &  1.05   &  9.85 &  0.200 &  45\\
$R_P$ ($R_J$) &  0.95   &  1.50 &  0.050 &  12\\
$T_0$         &  0.00   &  0.95 &  0.050 &  20\\
$\cos i$      &  0.0125 &  0.9875 &  0.025 &  40\\
\enddata
\label{tab:pars}
\end{deluxetable}

\begin{deluxetable}{crrrrrrrr}
\tabletypesize{\footnotesize}
\tablewidth{0pt}
\tablecaption{Artificial transit test statistics}
\tablehead{
\colhead{test} &
\multicolumn{2}{c}{all transits} &
\multicolumn{2}{c}{model} &
\multicolumn{2}{c}{marginal} &
\multicolumn{2}{c}{firm}\\
\colhead{type} &
\colhead{N}&\colhead{\%} &
\colhead{N}&\colhead{\%} &
\colhead{N}&\colhead{\%} &
\colhead{N}&\colhead{\%}}
\startdata
 1 & 40408 &    9.4 & 28696 &   71.0 &  7734 &   19.1 &  2818 &    7.0\\
 2 & 40408 &    9.4 & 28696 &   71.0 &  7493 &   18.5 &  2674 &    6.6\\
 3 & 40408 &    9.4 & 28696 &   71.0 &  4740 &   11.7 &  1045 &    2.6\\
\enddata
\label{tab:art}
\end{deluxetable}

\begin{deluxetable}{rrrrrrrrcc}
\tabletypesize{\footnotesize}
\tablewidth{0pc}
\tablecaption{Eclipsing binaries in NGC 2158}
\tablehead{\colhead{ID} & \colhead{$\alpha_{2000}$ [h]} &
\colhead{$\delta_{2000}$ [$\circ$]} &\colhead{P [d]} & \colhead{$R_{max}$} &
\colhead{$V_{max}$} &\colhead{$A_R$} & \colhead{$A_V$}&
\colhead{$P_1$} &\colhead{$P_2$}}
\startdata
 V01 &  6 07 32.6 & 23 49 21.0 &  0.2353 & 17.659 & 18.497  &  0.080 &  0.117 & \nodata & \nodata \\
 V58 &  6 07 47.8 & 23 51 16.0 &  0.2553 & 19.870 & 20.547  &  0.291 &  0.454 & \nodata & \nodata \\
 V59 &  6 07 23.9 & 24 04 57.5 &  0.3144 & 16.430 & 16.932  &  0.055 &  0.056 & \nodata & \nodata \\
 V02 &  6 07 23.3 & 24 06 12.3 &  0.3385 & 17.340 & 17.872  &  0.240 &  0.241 & \nodata & \nodata \\
 V60 &  6 07 38.0 & 24 07 24.9 &  0.3413 & 16.972 & 17.564  &  0.238 &  0.269 & 63 & 77  \\
 V03 &  6 07 04.9 & 23 48 53.3 &  0.3522 & 18.396 & 18.993  &  0.680 &  0.737 & \nodata & \nodata \\
 V04 &  6 07 17.5 & 24 04 45.7 &  0.3555 & 17.749 & 18.376  &  0.219 &  0.269 & \nodata & \nodata \\
 V05 &  6 07 40.6 & 24 05 03.6 &  0.3635 & 17.623 & 18.197  &  0.395 &  0.416 & \nodata & \nodata \\
 V61 &  6 06 48.3 & 23 52 44.0 &  0.3914 & 15.144 & 15.542  &  0.195 &  0.184 & \nodata & \nodata \\
 V07 &  6 06 37.0 & 23 50 41.1 &  0.5075 & 17.043 & 17.543  &  0.037 &  0.058 & \nodata & \nodata \\
 V08 &  6 06 44.8 & 24 06 57.8 &  0.5466 & 17.632 & 18.561  &  0.075 &  0.164 & \nodata & \nodata \\
 V09 &  6 07 29.4 & 24 10 06.3 &  0.6793 & 18.852 & 19.502  &  0.686 &  0.680 & \nodata & \nodata \\
 V10 &  6 07 34.9 & 24 04 25.6 &  0.8602 & 18.740 & 19.416  &  0.346 &  0.370 & \nodata & \nodata \\
 V17 &  6 07 31.9 & 24 06 00.2 &  0.8633 & 15.682 & 16.173  &  0.025 &  0.039 & \nodata & \nodata \\
 V62 &  6 07 32.8 & 24 07 26.6 &  0.8706 & 17.202 & 17.742  &  0.046 &  0.069 & \nodata & \nodata \\
 V11 &  6 07 21.4 & 24 05 39.7 &  0.9105 & 17.066 & 17.627  &  0.155 &  0.139 & \nodata & \nodata \\
 V19 &  6 06 20.5 & 24 04 47.8 &  0.9661 & 17.414 & 18.388  &  0.066 &  0.110 & \nodata & \nodata \\
 V12 &  6 07 18.7 & 24 06 50.0 &  1.0573 & 16.610 & 17.138  &  0.043 &  0.052 & 71 & 78  \\
 V63 &  6 07 37.5 & 23 55 57.3 &  1.0819 & 15.111 & 15.368  &  0.016 &  0.018 &  8 & 43  \\
 V13 &  6 07 28.1 & 24 06 35.3 &  1.2033 & 15.882 & 16.374  &  0.086 &  0.084 & \nodata & \nodata \\
 V14 &  6 07 27.2 & 23 52 24.4 &  1.2095 & 16.544 & 17.198  &  0.100 &  0.124 &  1 & 28  \\
 V15 &  6 07 36.1 & 23 48 24.0 &  1.5588 & 16.088 & 16.439  &  0.028 &  0.039 & \nodata & \nodata \\
 V16 &  6 06 36.6 & 24 03 29.1 &  1.5898 & 17.742 & 18.355  &  0.057 &  0.100 & \nodata & \nodata \\
 V64 &  6 07 17.8 & 24 05 32.5 &  1.7960 & 17.651 & 18.314  &  0.053 &  \nodata & \nodata & \nodata \\
 V65 &  6 07 39.3 & 24 02 16.0 &  1.8356 & 15.426 & 15.944  &  0.026 &  0.030 & 60 & 82  \\
 V18 &  6 07 31.0 & 24 05 50.4 &  1.9013 & 16.534 & 16.994  &  0.043 &  0.051 & \nodata & \nodata \\
 V66 &  6 07 31.4 & 24 05 18.4 &  2.0214 & 15.243 & 15.533  &  0.060 &  0.044 & \nodata & \nodata \\
 V20 &  6 07 20.6 & 24 06 01.2 &  2.0611 & 16.420 & 16.947  &  0.100 &  0.111 & \nodata & \nodata \\
 V21 &  6 07 36.3 & 23 56 21.3 &  2.0638 & 16.777 & 17.195  &  0.039 &  0.049 & 10 & 47  \\
 V22 &  6 07 37.7 & 24 07 40.2 &  2.1307 & 15.243 & 15.762  &  0.016 &  0.021 & 72 & 90  \\
 V23 &  6 07 03.5 & 24 01 45.1 &  2.2597 & 18.290 & 19.289  &  0.137 &  0.224 & \nodata & \nodata \\
 V24 &  6 07 43.4 & 24 06 22.8 &  2.3002 & 16.718 & 17.217  &  0.043 &  0.063 & 59 & 79  \\
 V67 &  6 07 24.1 & 24 06 22.7 &  2.3370 & 17.586 & 18.084  &  0.206 &  0.204 & \nodata & \nodata \\
 V25 &  6 07 19.9 & 24 06 24.4 &  2.3677 & 16.358 & 16.842  &  0.038 &  0.042 & \nodata & \nodata \\
 V68 &  6 07 27.6 & 24 10 48.8 &  2.4795 & 18.439 & 19.054  &  0.259 &  0.366 & \nodata & \nodata \\
 V27 &  6 07 30.2 & 24 07 50.5 &  2.4954 & 16.827 & 17.283  &  0.043 &  0.076 & \nodata & \nodata \\
 V26 &  6 07 24.1 & 24 07 53.8 &  2.6735 & 16.808 & 17.299  &  0.047 &  0.085 & \nodata & \nodata \\
 V34 &  6 07 19.9 & 23 49 51.4 &  2.8640 & 15.789 & 15.989  &  0.150 &  0.164 & \nodata & \nodata \\
 V28 &  6 06 50.0 & 24 08 26.0 &  2.9255 & 17.290 & 17.912  &  0.154 &  0.166 & 15 & 59  \\
 V29 &  6 06 46.2 & 23 49 41.3 &  3.1572 & 18.457 & 19.028  &  0.451 &  0.456 & \nodata & \nodata \\
 V30 &  6 07 14.9 & 24 09 40.8 &  3.3461 & 18.924 & 19.572  &  0.361 &  0.312 & \nodata & \nodata \\
 V69 &  6 07 22.2 & 24 06 15.4 &  3.7788 & 16.202 & 16.640  &  0.164 &  0.185 & \nodata & \nodata \\
 V31 &  6 06 49.0 & 24 01 43.7 &  3.7833 & 15.970 & 16.696  &  0.066 &  0.067 & 11 & 47  \\
 V32 &  6 07 27.9 & 24 06 22.7 &  4.5948 & 16.458 & 16.984  &  0.183 &  0.048 & \nodata & \nodata \\
 V33 &  6 07 33.3 & 24 04 34.4 &  4.7397 & 16.415 & 16.902  &  0.030 &  0.049 & \nodata & \nodata \\
 V70 &  6 06 36.9 & 24 02 05.9 &  6.2657 & 17.189 & 17.910  &  0.043 &  \nodata & \nodata & \nodata \\
 V55 &  6 07 25.8 & 24 05 45.7 &  8.1563 & 15.616 & 16.192  &  0.143 &  0.197 & \nodata & \nodata \\
\enddata
\tablecomments{Cluster core and corona membership probabilities $P_1$ and $P_2$ taken from K97.}
\label{tab:ecl}
\end{deluxetable}

\begin{deluxetable}{rrrrrrrccc}
\tabletypesize{\footnotesize}
\tablewidth{0pc}
\tablecaption{Other periodic variables in NGC 2158}
\tablehead{\colhead{ID} & \colhead{$\alpha_{2000}$ [h]} &
\colhead{$\delta_{2000}$ [$\circ$]} &\colhead{P [d]} & \colhead{$\langle R\rangle$} &
\colhead{$\langle V\rangle$} &\colhead{$A_R$} & \colhead{$A_V$}&
\colhead{$P_1$} &\colhead{$P_2$}}
\startdata
 V35 &  6 06 47.1 & 24 07 41.5 &  0.0981 & 15.820 & 16.351  &  0.008 &  0.014 & 13 & 58  \\
 V36 &  6 07 16.8 & 24 05 44.9 &  0.1078 & 15.987 & 16.473  &  0.022 &  0.033 & 83 & 90  \\
 V71 &  6 07 23.8 & 24 10 40.4 &  0.1112 & 16.210 & 16.675  &  0.003 &  0.005 & \nodata & \nodata \\
 V37 &  6 07 26.0 & 24 04 23.6 &  0.1113 & 15.794 & 16.236  &  0.037 &  0.048 & \nodata & \nodata \\
 V38 &  6 07 11.3 & 24 03 36.4 &  0.1205 & 15.619 & 16.053  &  0.005 &  0.007 &  3 &  4  \\
 V39 &  6 06 50.6 & 24 10 56.5 &  0.1220 & 15.211 & 15.648  &  0.018 &  0.024 &  4 & 23  \\
 V72 &  6 06 11.4 & 24 08 52.9 &  0.3084 & 18.187 & 19.282  &  0.020 &  0.021 & \nodata & \nodata \\
 V73 &  6 06 50.6 & 24 09 58.3 &  0.3729 & 18.383 & 19.549  &  0.020 &  0.019 & \nodata & \nodata \\
 V74 &  6 06 27.9 & 24 03 47.4 &  0.6762 & 19.560 & 20.687  &  0.064 &  0.053 & \nodata & \nodata \\
 V75 &  6 06 32.1 & 24 08 51.5 &  0.6778 & 15.844 & 16.195  &  0.003 &  0.003 &  2 & 37  \\
 V76 &  6 07 23.1 & 24 10 00.5 &  0.7981 & 16.787 & 17.263  &  0.005 &  0.005 & 54 & 79  \\
 V77 &  6 07 25.4 & 24 09 21.6 &  1.1442 & 16.657 & 17.116  &  0.007 &  0.008 & \nodata & \nodata \\
 V78 &  6 07 37.6 & 24 05 57.3 &  1.2027 & 16.786 & 17.277  &  0.004 &  0.005 & \nodata & \nodata \\
 V79 &  6 07 42.2 & 23 50 39.0 &  1.3118 & 18.337 & 19.026  &  0.028 &  0.028 & \nodata & \nodata \\
 V40 &  6 07 05.6 & 24 07 06.5 &  1.3592 & 18.178 & 19.033  &  0.018 &  0.022 & \nodata & \nodata \\
 V80 &  6 07 32.5 & 24 10 09.7 &  1.4727 & 16.799 & 17.229  &  0.007 &  0.009 & \nodata & \nodata \\
 V41 &  6 07 04.2 & 23 58 09.5 &  1.4949 & 17.079 & 17.795  &  0.026 &  0.030 &  7 & 28  \\
 V42 &  6 07 36.3 & 24 02 07.4 &  1.6210 & 17.425 & 17.911  &  0.031 &  0.037 & 67 & 88  \\
 V43 &  6 07 49.9 & 24 09 45.5 &  2.6060 & 16.588 & 17.056  &  0.008 &  0.007 & \nodata & \nodata \\
 V44 &  6 07 05.8 & 24 08 51.3 &  2.8294 & 16.684 & 17.391  &  0.016 &  0.021 & 44 & 80  \\
 V81 &  6 07 40.6 & 24 05 50.0 &  2.9942 & 15.891 & 16.196  &  0.011 &  0.009 & 19 & 23  \\
 V45 &  6 06 27.8 & 23 49 58.6 &  5.0129 & 18.937 & 19.845  &  0.036 &  0.050 & \nodata & \nodata \\
 V82 &  6 06 24.2 & 24 06 43.5 &  5.3317 & 15.476 & 16.109  &  0.003 &  0.004 & \nodata & \nodata \\
 V46 &  6 07 09.7 & 24 06 50.2 &  5.4488 & 17.378 & 18.153  &  0.014 &  0.024 & \nodata & \nodata \\
 V47 &  6 06 40.4 & 24 06 34.3 &  5.4962 & 17.255 & 17.931  &  0.009 &  0.011 &  7 & 51  \\
 V83 &  6 07 49.4 & 24 09 13.2 &  5.8010 & 15.529 & 16.050  &  0.009 &  0.012 & \nodata & \nodata \\
 V54 &  6 07 07.0 & 24 05 25.3 &  6.3310 & 15.373 & 15.891  &  0.009 &  0.009 & \nodata & \nodata \\
 V48 &  6 07 06.2 & 24 02 10.1 &  6.4713 & 16.422 & 17.115  &  0.025 &  0.031 & 41 & 69  \\
 V49 &  6 07 10.2 & 24 10 19.4 &  6.5614 & 16.520 & 17.231  &  0.024 &  0.032 & 26 & 51  \\
 V84 &  6 07 31.8 & 23 48 28.9 &  7.0849 & 17.284 & 18.303  &  0.010 &  0.015 & \nodata & \nodata \\
 V85 &  6 06 22.3 & 24 07 09.1 &  7.2146 & 15.630 & 16.295  &  0.005 &  0.006 & \nodata & \nodata \\
 V50 &  6 06 43.0 & 23 55 15.1 &  7.8975 & 16.266 & 16.803  &  0.011 &  0.014 &  0 & 24  \\
 V86 &  6 07 22.7 & 23 48 19.0 &  9.1927 & 17.706 & 18.749  &  0.020 &  0.028 & \nodata & \nodata \\
 V87 &  6 06 43.9 & 23 54 04.9 &  9.3788 & 16.509 & 16.953  &  0.006 &  0.012 & \nodata & \nodata \\
 V53 &  6 07 49.0 & 24 02 02.6 &  9.8949 & 17.023 & 17.849  &  0.014 &  0.016 & 34 & 65  \\
 V88 &  6 07 47.5 & 24 04 42.4 &  9.9873 & 16.907 & 17.715  &  0.008 &  0.010 & 25 & 38  \\
 V89 &  6 07 36.4 & 24 05 42.3 & 12.5495 & 15.673 & 16.299  &  0.008 &  0.008 & \nodata & \nodata \\
 V90 &  6 07 36.2 & 24 05 25.4 & 12.8136 & 15.924 & 16.580  &  0.005 &  0.008 & \nodata & \nodata \\
 V52 &  6 06 24.0 & 23 54 25.6 & 16.0765 & 16.062 & 16.523  &  0.007 &  0.012 & \nodata & \nodata \\
 V91 &  6 07 35.2 & 23 50 38.2 & 20.7403 & 17.337 & 18.007  &  0.009 &  0.008 & \nodata & \nodata \\
 V92 &  6 07 24.7 & 24 03 58.9 & 30.4984 & 16.333 & 17.075  &  0.011 &  0.014 & \nodata & \nodata \\
 V93 &  6 06 36.3 & 24 04 17.9 & 36.8876 & 17.759 & 19.219  &  0.040 &  0.060 & \nodata & \nodata \\
 V56 &  6 06 17.2 & 24 03 39.7 & 63.4571 & 16.746 & 17.305  &  0.029 &  0.016 & \nodata & \nodata \\
\enddata
\tablecomments{Cluster core and corona membership probabilities $P_1$ and $P_2$ taken from K97.}
\label{tab:pul}
\end{deluxetable}

\begin{deluxetable}{ccccccccc}
\tabletypesize{\footnotesize}
\tablewidth{0pc}
\tablecaption{Properties of $\delta$ Scuti stars in NGC 2158}
\tablehead{\colhead{ID} & \colhead{P$_1$ [d]}& \colhead{P$_2$ [d]} 
&\colhead{P$_3$ [d]} &\colhead{P$_4$ [d]} &\colhead{P$_5$ [d]} &
\colhead{P$_6$ [d]} & \colhead{$\Delta$(m-M)$_0^{min}$} & 
\colhead{$\Delta$(m-M)$_0^{max}$}}
\startdata
V35& 0.098051& 0.104421& 0.090706& 0.084631&0.114466&0.084468& -0.055& -0.398\\
V36& 0.107822& 0.111340& 0.114872& \nodata &\nodata &\nodata & 0.119 & 0.221\\
V37& 0.111281& 0.106033& 0.113535& \nodata &\nodata &\nodata & -0.035& -0.145\\
V38& 0.120463& 0.074820& 0.080215& \nodata &\nodata &\nodata & -0.122& -0.892\\
V39& 0.121997& \nodata & \nodata & \nodata &\nodata &\nodata & -0.506& -0.506\\
V71& 0.111217& 0.097834& 0.084016& 0.088923&\nodata &\nodata & 0.009 & 0.371\\
\enddata
\label{tab:mm}
\end{deluxetable}

\begin{deluxetable}{rrrlccccccc}
\tabletypesize{\footnotesize}
\tablewidth{0pc}
\tablecaption{Miscellaneous variables in NGC 2158}
\tablehead{\colhead{ID} & \colhead{$\alpha_{2000}$ [h]} &
\colhead{$\delta_{2000}$ [$\circ$]} & \colhead{$R_{max}$} &
\colhead{$V_{max}$} &\colhead{$A_R$} &\colhead{$A_V$}&
\colhead{$P_1$} &\colhead{$P_2$}}
\startdata
 V57 &  6 07 33.8 & 24 07 55.2 & 18.322 & 18.626  &  3.495 &  4.317 & \nodata & \nodata \\
 V94 &  6 07 27.2 & 24 04 35.5 & 15.408 & 16.197  &  0.057 &  0.066 & \nodata & \nodata \\
 V95 &  6 07 18.8 & 24 01 40.1 & 15.922 & 16.703  &  0.039 &  0.045 & 53 & 70  \\
 V96 &  6 07 32.0 & 24 06 33.7 & 16.363 & 16.865  &  0.139 &  0.149 & \nodata & \nodata \\
 V97 &  6 07 17.3 & 24 02 46.5 & 16.824 & 17.350  &  0.116 &  0.142 & \nodata & \nodata \\
\enddata
\label{tab:misc}
\end{deluxetable}


\begin{references}
\reference{} Alard, C.\ 2000, A\&AS, 144, 363
\reference{} Alard, C., Lupton, R.\ 1998, ApJ, 503, 325
\reference{} Albrow, M.~D., Gilliland, R.~L., Brown, T.~M., Edmonds,  
             P.~D., Guhathakurta, P., \& Sarajedini, A.\ 2001, \apj,
             559, 1060
\reference{} Alonso, R., Belmonte, J.~A., \& Brown, T.\ 2003, \apss, 284, 13
\reference{} Alonso, R., et al.\ 2004, \apjl, 613, L153
\reference{} Arentoft, T., Bouzid, M.~Y., Sterken, C., Freyhammer,
             L.~M., \& Frandsen, S.\ 2005, \pasp, 117, 601
\reference{} Bakos, G., Noyes, R.~W., Kov{\' a}cs, G., Stanek, K.~Z.,
             Sasselov, D.~D., \& Domsa, I.\ 2004, \pasp, 116, 266
\reference{} Baraffe, I., Chabrier, G., Allard, F., \& Hauschildt,
             P.~H.\ 2002, \aap, 382, 563
\reference{} Brown, T.~M., Charbonneau, D., Gilliland, R.~L., Noyes,
             R.~W., \& Burrows, A.\ 2001, \apj, 552, 699
\reference{} Bruntt, H., Grundahl, F., Tingley, B., Frandsen, S.,
             Stetson, P.~B., \& Thomsen, B.\ 2003, \aap, 410, 323
\reference{} Carraro, G., Girardi, L., \& Marigo, P.\ 2002, \mnras,
             332, 705 (Ca02)
\reference{} Charbonneau, D., et al.\ 2005, \apj, submitted
             (astro-ph/0508051)
\reference{} Christian, C.~A., Heasley, J.~N., \& Janes, K.~A.\ 1985,
             \apj, 299, 683
\reference{} Cutri, R. M., et al. 2003, The 2MASS All-Sky Catalog of
             Point Sources (Pasadena: IPAC)
\reference{} Girardi, L., Bressan, A., Bertelli, G., \& Chiosi, C.\
             2000, \aaps, 141, 371
\reference{} Hebb, L., Wyse, R.~F.~G., \& Gilmore, G.\ 2004, \aj, 128,
              2881
\reference{} Hood, B., et al.\ 2005, \mnras, 360, 791 
\reference{} Kafka, S., \& Honeycutt, R.~K.\ 2005, \aj, 130, 742
\reference{} Kaluzny, J., Stanek, K.~Z., Garnavich, P.~M. \& Challis, P.\ 
             1997, ApJ, 491, 153
\reference{} Kane, S.~R., Lister, T.~A., Collier Cameron, A., Horne,
             K., James, D., Pollacco, D.~L., Street, R.~A., \&
             Tsapras, Y.\ 2005, \mnras, 362, 117
\reference{} Kharchenko, N., Andruk, V., \& Schilbach, E.\ 1997,
             Astronomische Nachrichten, 318, 253 (KAS97)
\reference{} Konacki, M., Torres, G., Jha, S., \& Sasselov, D.~D.\
             2003, \nat, 421, 507
\reference{} Kov{\' a}cs, G., Zucker, S., \& Mazeh, T.\ 2002, \aap,
             391, 369
\reference{} Mall{\' e}n-Ornelas, G., Seager, S., Yee, H.~K.~C.,
             Minniti, D., Gladders, M.~D., Mall{\' e}n-Fullerton,
             G.~M., \& Brown, T.~M.\ 2003, \apj, 582, 1123
\reference{} Marshall, J.~L., Burke, C.~J., DePoy, D.~L., Gould, A.,
             \& Kollmeier, J.~A.\ 2005, \aj, 130, 1916
\reference{} Mathieu, R.~D., van den Berg, M., Torres, G., Latham, D.,
             Verbunt, F., \& Stassun, K.\ 2003, \aj, 125, 246
\reference{} Mochejska, B.~J., Stanek, K.~Z., Sasselov, D.~D.,
             \& Szentgyorgyi, A.~H.\ 2002, \aj, 123, 3460 (Paper~I) 
\reference{} Mochejska, B.~J., Stanek, K.~Z., Sasselov, D.~D.,
             Szentgyorgyi, A.~H., Westover, M., \& Winn, J.~N.\ 2004,
             \aj, 128, 312 (Paper~II)
\reference{} Mochejska, B.~J., et al.\ 2005, \aj, 129, 2856 (Paper~III)
\reference{} Petersen, J.~O.~\& Christensen-Dalsgaard, J.\ 1999, \aap, 352,
             547 
\reference{} Pinfield, D.~J., Jones, H.~R.~A., \& Steele, I.~A.\ 2005,
             \pasp, 117, 173
\reference{} Pont, F., Melo, C.~H.~F., Bouchy, F., Udry, S., Queloz,
             D., Mayor, M., \& Santos, N.~C.\ 2005, \aap, 433, L21
\reference{} Santos, N.~C., Israelian, G., \& Mayor, M.\ 2004, \aap,
             415, 1153
\reference{} Stetson, P.~B.\ 1987, PASP, 99, 191
\reference{} Tamuz, O., Mazeh, T., \& Zucker, S.\ 2005, \mnras, 356, 1466 
\reference{} Udalski, A., Pietrzynski, G., Szymanski, M., Kubiak, M.,
             Zebrun, K., Soszynski, I., Szewczyk, O., \& Wyrzykowski,
             L.\ 2003, Acta Astronomica, 53, 133
\reference{} Udalski, A., Szewczyk, O., Zebrun, K., Pietrzynski, G.,
             Szymanski, M., Kubiak, M., Soszynski, I., \& Wyrzykowski,
             L.\ 2002c, Acta Astronomica, 52, 317
\reference{} Udalski, A., Zebrun, K., Szymanski, M., Kubiak, M.,
             Soszynski, I., Szewczyk, O., Wyrzykowski, L., \&
             Pietrzynski, G.\ 2002b, Acta Astronomica, 52, 115
\reference{} Udalski, A., et al.\ 2002a, Acta Astronomica, 52, 1
\reference{} von Braun, K., Lee, B.~L., Mall{\' e}n-Ornelas, G., Yee,
             H.~K.~C., Seager, S., \& Gladders, M.~D.\ 2004, AIP
             Conf.~Proc.~713: The Search for Other Worlds, 713, 181
\reference{} Warner, B.\ 1995, Cambridge Astrophysics Series,
             Cambridge, New York: Cambridge University Press
\end{references}
\end{document}